\documentclass[12pt]{article} 
\usepackage[a4paper, margin=1in]{geometry} 
\usepackage{graphicx}   
\usepackage{amsmath, amssymb} 
\usepackage{xcolor}     
\usepackage{soul}       
\sethlcolor{white}      
\usepackage{booktabs}   
\usepackage{multirow}   
\usepackage{changepage}  
\newlength{\extralength}       
\usepackage[colorlinks=true, linkcolor=blue, urlcolor=blue]{hyperref}

\title{Semi-Automatic Flute Robot and Its Acoustic Sensing}

\author{
Hikari Kuriyama\thanks{Graduate School of Science and Technology, Kumamoto University, Kumamoto 860-8555, Japan; \texttt{kuriyama@navi.cs.kumamoto-u.ac.jp}} \and
Hiroaki Sonoda\thanks{Graduate School of Science and Technology, Kumamoto University, Kumamoto 860-8555, Japan; \texttt{sonoda@navi.cs.kumamoto-u.ac.jp}} \and
Kouki Tomiyoshi\thanks{Graduate School of Science and Technology, Kumamoto University, Kumamoto 860-8555, Japan; \texttt{tomiyoshi@navi.cs.kumamoto-u.ac.jp}} \and
Gou Koutaki\thanks{Department of Computer Science and Electrical Engineering, Kumamoto University, Kumamoto 860-8555, Japan; \texttt{koutaki@cs.kumamoto-u.ac.jp}} 
}

\date{} 

\begin{document}

\maketitle

\begin{abstract}
Flute performance requires mastery of complex fingering combinations and register-dependent embouchure control, particularly jet offset adjustment for low-register production. Existing haptic and semi-automated systems do not address both aspects simultaneously through mechanical actuation. To our knowledge, no prior system fully automates fingering while mechanically assisting low-register tone production without requiring embouchure control.
We developed a semi-automatic flute robot with an automatic fingering mechanism: fourteen servo motors actuate all keys via wire-based and rack-and-pinion drives in response to MIDI input, enabling performers to produce complete musical pieces through airflow alone. A jet offset assist mechanism rotates the head joint by a calibrated $22^\circ$ during low-register passages, shifting the jet offset toward a low-register configuration without modifying the instrument or embouchure.
Fundamental frequency estimation confirmed correct pitch production across the chromatic range (C4--C7) and during musical performance. All key and lever movements were completed within 77.50~ms, corresponding to tempo capacity exceeding standard requirements. Harmonic analysis ($\Delta\mathrm{SPL} = \mathrm{SPL}_2 - \mathrm{SPL}_3$) showed a consistent increase in $\Delta$SPL for all low-register notes when activated, consistent with the intended jet offset shift. Head joint rotation completed within 40.00~ms.
These results demonstrate mechanical feasibility of integrating automated fingering and register-dependent jet offset assistance under controlled conditions.
\end{abstract}

\textbf{Keywords:} flute; robotics; semi-automatic; performance support; automatic fingering; jet offset;


\section{Introduction}

Learning to play the flute presents two distinct technical challenges that are 
difficult to address simultaneously. 
The first is fingering: the flute requires more than 25 different fingering combinations using nearly all fingers, and mastering them demands substantial practice.
The second is register control: in the low and middle registers, many notes share the same fingering, and pitch is differentiated primarily through breath speed and \textit{jet offset} — the extent to which the center of the air jet is biased toward the inside of the flute from the edge of the embouchure hole~\cite{ando_book}. 
Experienced players control jet offset through embouchure adjustment, but this is extremely challenging for beginners, and is a primary reason why low-register note production is particularly difficult at early stages of learning.

Systems that automate musical instrument performance entirely have been reported across a range of instruments~\cite{Solis_flute_sax,Solis_flute,Lin_sax,Indreica_piano,Jo_violin,
Toyota_violin,Murphy_guitar,Jiayin_bamboo,Fei_daru,Kato_taiflute}. 
These fully automated robots are designed for performance analysis or musical appreciation and are not intended to support human performers directly.

A different category of systems automates part of the performance while leaving the remaining actions to the 
human~\cite{Koutaki_sax,Tsurumi,Huijiang_piano}. 
These systems are designed to support beginners and individuals with physical constraints, or are developed for other primary purposes with such applications envisioned as extensions. They are expected to facilitate musical instrument performance.
Among performance-support systems specifically, haptic approaches such as MoveMe~\cite{Fujii_MoveMe} and ShIFT~\cite{Xia_semi} guide the performer's 
hand and finger movements through tactile feedback. 
While effective for learning support, these systems still require the performer to execute fingering motions, which limits their utility when the goal is performing a complete musical piece with minimal physical constraint.
Furthermore, no existing system specifically addresses the difficulty of low-register tone production through a mechanical assist.
A detailed review of related work is provided in Section~\ref{sec:related}.

This study proposes a \textit{semi-automatic flute robot} consisting of two mechanisms mountable on a standard flute. 
The \textit{automatic fingering mechanism} uses servo motors to press keys in response to MIDI~\cite{MIDI} input, allowing users who can produce basic flute sounds to perform musical pieces by blowing, without managing fingering.
The \textit{jet offset assist mechanism} rotates the head joint by a predetermined angle during low-register passages, shifting the jet offset toward a value suitable for low-register production.
The scope of this study is limited to evaluating the mechanical feasibility of both mechanisms; user-centered evaluation is left as future work.

The automatic fingering mechanism is validated through fundamental frequency estimation across the full pitch range and during musical piece performance, movement time measurement.
The jet offset assist mechanism is evaluated through harmonic analysis comparing sound pressure levels of the second and third harmonics with and without the mechanism, and through head joint movement time measurement.

The main contributions of this study are as follows:

\begin{itemize}
    \item We develop a semi-automatic flute robot that combines an automatic 
    fingering mechanism and a jet offset assist mechanism, enabling users who 
    can produce basic flute tones to perform musical pieces by blowing into 
    the instrument.
    
    \item We propose and mechanically validate a novel approach to low-register 
    assistance by rotating the head joint to adjust the jet offset toward a 
    value suitable for the low register.
    
    \item Through mechanical evaluation, we characterize the operational 
    behavior, performance limits, and acoustic properties of both mechanisms, 
    identifying motor noise reduction as the primary design priority for 
    future development.
\end{itemize}

\begin{figure}[tb]
    \centering 
    \includegraphics[width=0.7\linewidth]{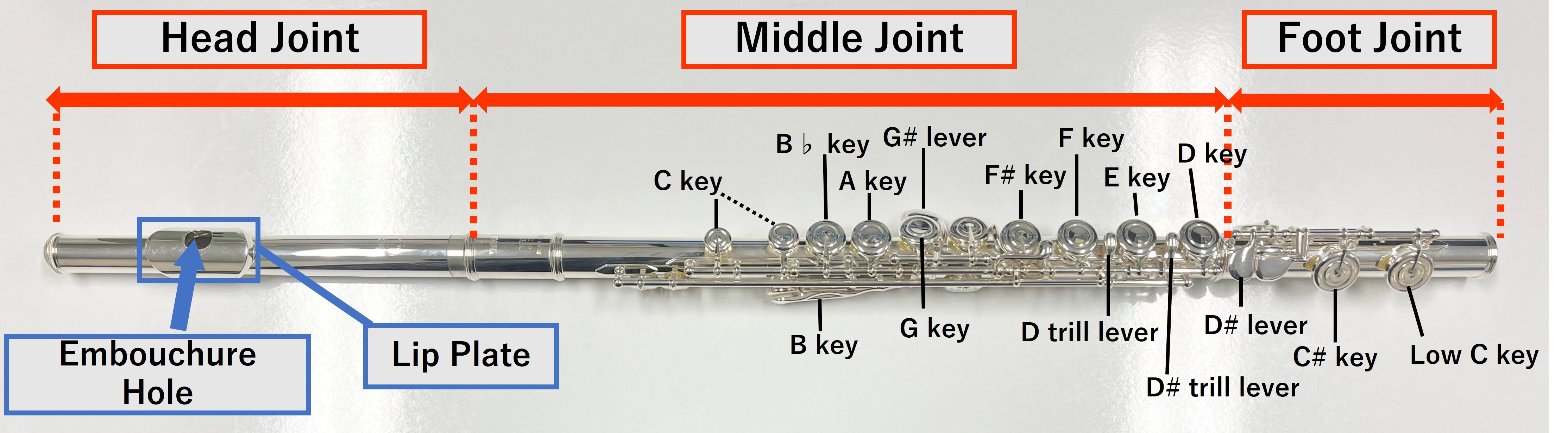}
    \caption{The flute has three main sections: the head joint, middle joint, and foot joint. The head joint features a lip plate with a central embouchure hole. The body and foot joint are equipped with multiple keys and levers; pressing a key closes its tone hole, while pressing a lever opens the corresponding key.}
    \label{fig:flute}
\end{figure}

\section{Related Work}
\label{sec:related}

This section reviews work directly relevant to positioning the present study: fully automated flute robots, semi-automated and haptic performance-support systems, and wind-instrument assistive devices. 

\subsection{Fully Automated Flute Robots}
Two lines of fully automated flute research are directly relevant as a 
baseline against which semi-automatic approaches are motivated.

Solis et al. developed an anthropomorphic robot capable of fully automated flute and saxophone performance~\cite{Solis_flute_sax, Solis_flute}. 
The system reproduces 41 degrees of freedom, mechanically replicating the lips, lungs, fingers, and tonguing mechanism, enabling autonomous performance without any human involvement.
Li et al. developed a fully automated robot for Chinese bamboo flute, a transverse flute with structural similarities to the Western flute~\cite{Jiayin_bamboo}. 
The system uses pneumatic actuated fingers and an air-blowing nozzle, and performs autonomously.
Both systems are designed for performance analysis or musical appreciation and are not intended to directly assist or interact with a human performer during playing.

Hurtado et al. proposed a robotic system aimed at creating an experimental environment that enables precise control and measurement of the strategies used by flutists while controlling their instruments\cite{Hurtado1,Hurtado2}.
The system independently controls three acoustically critical parameters: air jet length, jet angle, and jet offset relative to the labium. 
While this work is the most directly relevant prior study of mechanical jet offset control, it is an instrumented research platform rather than a performance-support system, and it does not address the needs of a human performer.

\subsection{Semi-Automated and Haptic Performance-Support Systems}

A distinct category of systems automates part of the performance while 
leaving the remaining actions to the human performer.

Fujii et al. developed MoveMe, a musical performance support system that provides three-dimensional haptic guidance, physically leading the performer’s hand movements as if another person were guiding them, thereby supporting beginner musicians~\cite{Fujii_MoveMe}.
Xia et al. developed ShIFT, a semi-haptic interface for tin whistle performance that provides tactile guidance for fingering motions while leaving breath control entirely to the performer~\cite{Xia_semi}. 
ShIFT is the most structurally comparable prior system to the present work: it targets the same problem space (reducing the fingering burden on beginners) and adopts a semi-automated philosophy in which only part of the performance is assisted.
Both MoveMe and ShIFT are characterized by guiding the performer's physical movements through feedback, but in both cases the performer must still execute the fingering motions themselves.

In the domain of semi-automated performance more broadly, Koutaki and Hamanaka developed an automatic fingering system for saxophone quartet~\cite{Koutaki_sax}, and Tsurumi et al. proposed a detachable semi-automatic support robot for guitar~\cite{Tsurumi}. 
These systems automate key or string actuation mechanically, allowing the performer to focus on breath or strumming rather than fingering precision. 
Wang et al. proposed a theoretical framework for human–robot cooperative piano playing based on non-verbal cues~\cite{Huijiang_piano}.
These works collectively establish that mechanical actuation of instrument controls, rather than haptic guidance, is a viable paradigm for performance support.

\subsection{Flute-Specific and Wind Instrument Support Systems}

Within flute-specific support, two systems address aspects of the sound-production problem.

Heller et al. developed an augmented flute that measures and visualizes embouchure parameters to assist beginners who would otherwise require an experienced teacher to identify and correct~\cite{Heller_2017}. 
The system makes embouchure parameters observable but does not modify them mechanically; correction remains the performer's responsibility.
Kuroda et al. proposed a machine-learning-based system that estimates flute control parameters (including embouchure and breath characteristics) from recorded audio, suggesting the potential for performance guidance without sensors~\cite{Kuroda_2022}.

For wind instruments more broadly, Kato et al. proposed a breathing assist device for saxophone players with respiratory difficulties~\cite{Kato_sax_assist1, Kato_sax_assist2}. 
This work provides a precedent for mechanical intervention in wind instrument performance to compensate for a specific physical limitation, an approach analogous in principle to the jet offset assist mechanism proposed here.

\subsection{Gap in Existing Research}

The review above reveals a consistent and specific gap. 
Fully automated flute robots (Solis, Hurtado, Li) demonstrate that mechanical systems can control all aspects of flute performance, but they are closed systems with no role for a human performer. 
Haptic and semi-haptic support systems (MoveMe, ShIFT) reduce the learning burden by guiding the performer's movements, but the performer must still execute every fingering motion; this limits their utility when performing varied or demanding musical pieces. 
Flute-specific assistive tools (Heller, Kuroda) address sound-production feedback but provide no mechanical intervention during performance.
Critically, no existing system simultaneously (i) automates fingering actuation mechanically so that the performer is entirely freed from finger-key coordination, while (ii) leaving sound production — breath control, articulation, and dynamics — under the performer's direct control. 
Furthermore, no prior system specifically addresses the mechanical difficulty of low-register tone production, which requires precise jet offset control that is well beyond the ability of most beginner players.

The present study addresses both gaps directly. 
The automatic fingering mechanism removes the fingering constraint entirely through mechanical key actuation, allowing the performer to focus on breath and sound production. 
The jet offset assist mechanism introduces, for the first time, a mechanical approach to low-register support by rotating the head joint to shift the jet offset toward a value suitable for the low register during performance. 
Together, these mechanisms define a new position in the design space of performance-support systems: mechanical actuation of fingering combined with performer-controlled sound production, extended to address a specific acoustic challenge that existing systems have not attempted to solve.

\section{Proposed system}
\label{sec:system}

\begin{figure}[tb]
    \centering
    
    \begin{minipage}{0.9\linewidth}
        \centering
        \includegraphics[width=\linewidth]{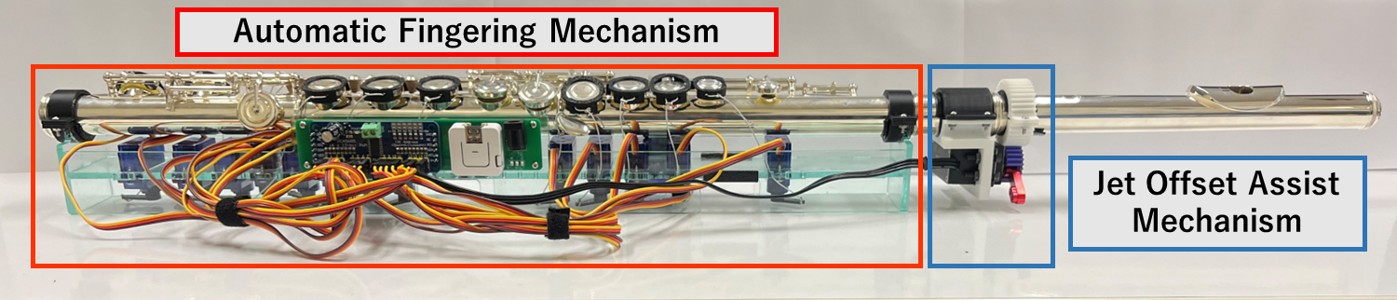}
        \small (a) 
    \end{minipage}
    
    \vspace{0.5em}
    
    \begin{minipage}{0.9\linewidth}
        \centering
        \includegraphics[width=\linewidth]{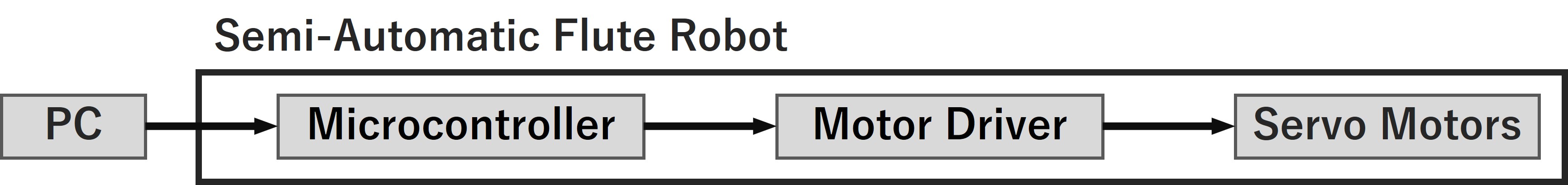}
        \small (b) 
    \end{minipage}
    
    \caption{Semi-automatic flute robot. (a) Overview of the system. Red highlights: automatic fingering mechanism (middle and foot joints). Blue highlights: jet offset assist mechanism (head and middle joints). (b) Control flow of the system. Both mechanisms are driven by servo motors controlled via MIDI messages from a PC.}
    \label{fig:flute_robot_all}
\end{figure}

The proposed semi-automatic flute robot consists of two independently operable mechanisms mounted on a standard covered-key flute (Figure~\ref{fig:flute_robot_all}): an \textit{automatic fingering mechanism} that actuates keys via servo motors in response to MIDI input, and a \textit{jet offset assist mechanism} that rotates the head joint to shift 
the jet offset toward a value suitable for low-register playing.
Both mechanisms share four design constraints that govern all subsequent design decisions: 
(i) the design must allow the maintenance of a normal flute playing posture,
(ii) the design must prioritize safety,
(iii) the structure must be non-destructive and detachable from the flute, and
(iv) the design must be lightweight.

\subsection{Automatic Fingering Mechanism}
\label{subsec:fingering}

\subsubsection{Design Rationale}

This study uses a covered-key flute, in which each tone hole is closed 
by pressing a lid-like key rather than requiring direct finger contact 
with the hole. 
This property makes full mechanical actuation of key closure possible 
without replicating the geometry of a human finger.

\begin{figure}[tb]
    \centering 
    \includegraphics[width=0.5\linewidth]{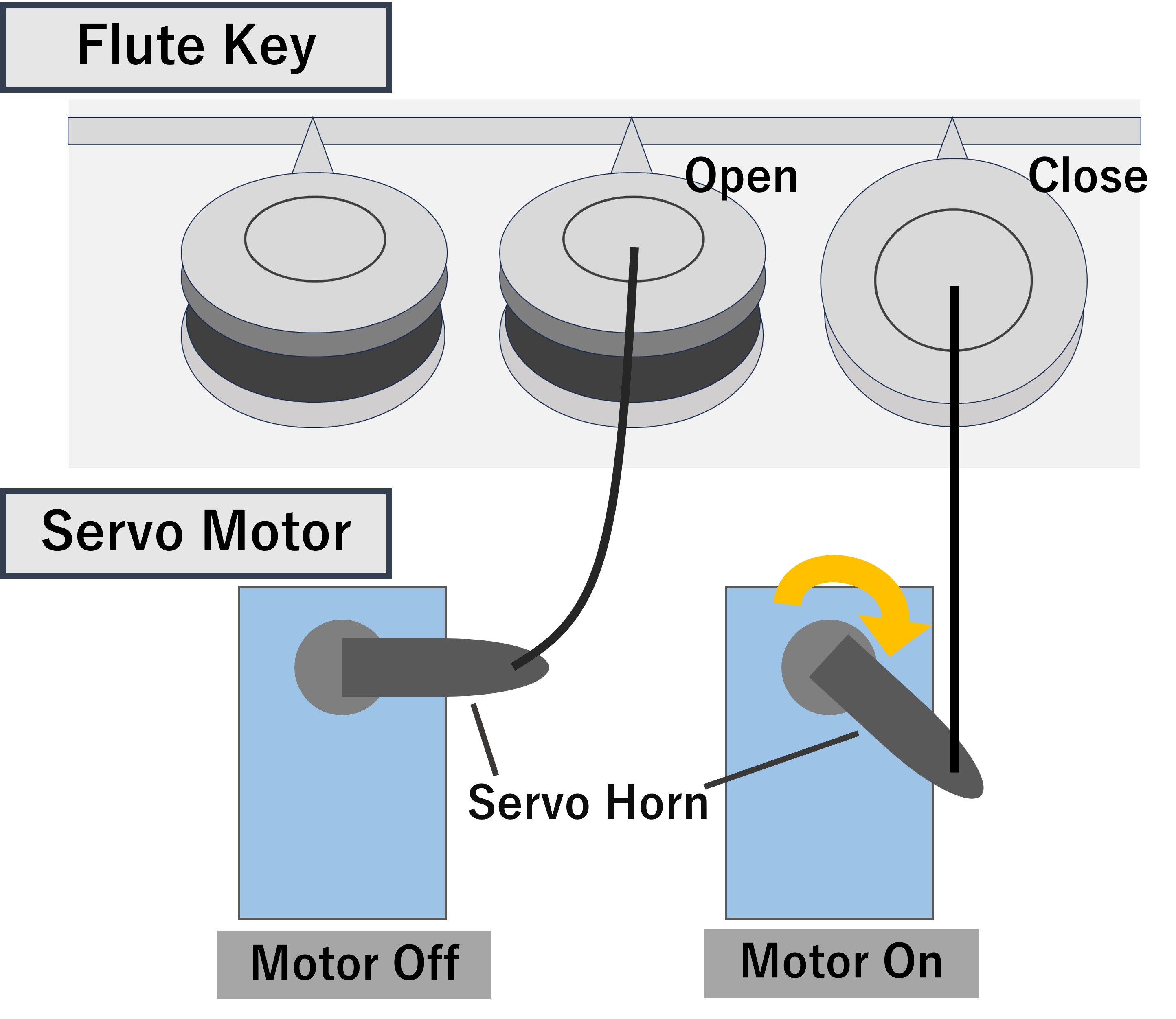}
    \caption{Schematic of wire-based key actuation. The wire connects the key cover to the servo horn; motor rotation pulls the wire and presses the key.}
    \label{fig:wire_mechanism}
\end{figure}

Pressing keys directly from above — as a human player does — would require a mechanism of impractical size and weight. 
Instead, a wire-based actuation method is adopted  (Figure~\ref{fig:wire_mechanism}): a wire attached to each key is pulled by servo motor rotation, pressing the key to close the tone hole. 
To satisfy the non-destructive and detachable constraint, wires are not fixed directly to the keys but are connected via removable key covers designed to apply force stably and uniformly to the key surface.

\subsubsection{Mechanism Overview}

\begin{figure}[tb]
 \centering
     \begin{minipage}{0.8\linewidth}
         \centering
         \includegraphics[width=\linewidth]{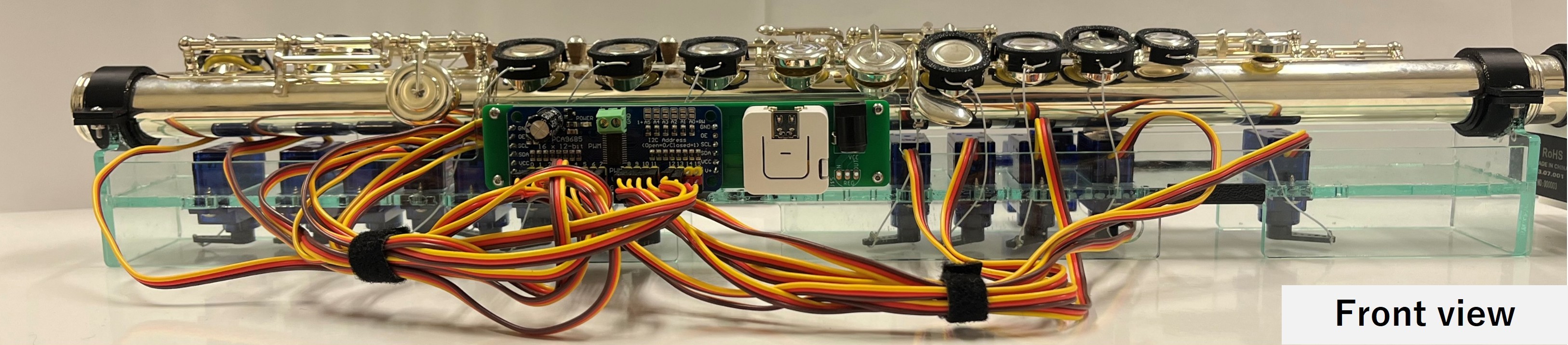}
         \small (a) 
     \end{minipage}
    
     \vspace{1mm}

     \begin{minipage}{0.8\linewidth}
         \centering
         \includegraphics[width=\linewidth]{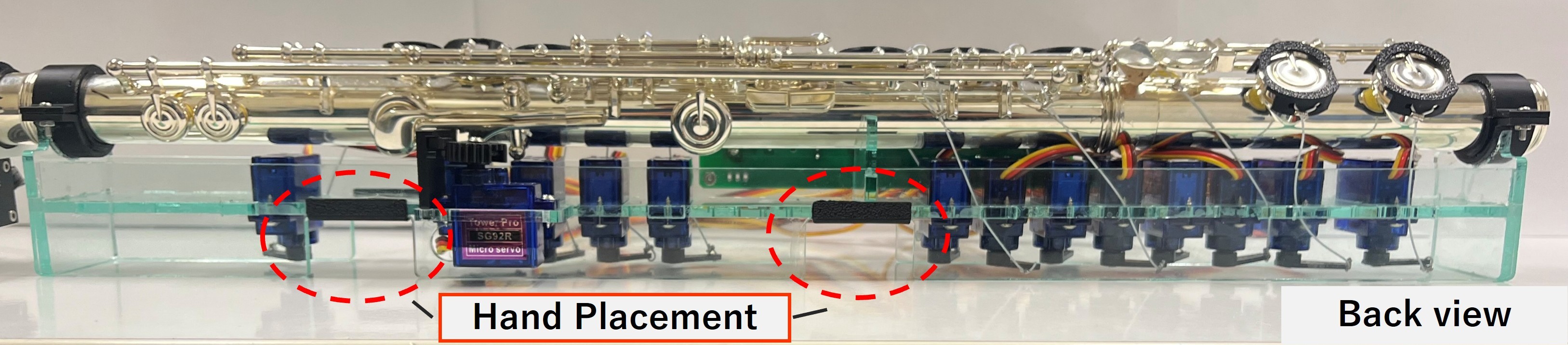}
         \small (b) 
     \end{minipage}
    \vspace{-2.5pt}
    \caption{Automatic fingering mechanism. (a) Front view: servo motors are mounted on an acrylic body, with the flute secured on top.
    (b) Rear view: a hand-placement grip is positioned to support a natural playing posture.}
    \label{fig:flute_robot_auto}
\end{figure}

The mechanism mounts on an acrylic body that houses fourteen SG92R servo motors (Tower Pro), each assigned to one key 
(Figure~\ref{fig:flute_robot_auto}).
The flute is secured to the body via 3D-printed fixtures with bolt-and-nut attachment; a silicone sheet between the fixture and the flute prevents slippage without damaging the instrument surface. 
A hand-placement grip positioned near the standard flute support point allows the player to hold the assembly while maintaining a natural playing posture.

Two actuation methods are used depending on key geometry:

\paragraph{\underline{Wire-Based Method}}

\begin{figure}[tb]
    \centering
    \begin{minipage}[b]{0.3\textwidth}
        \centering
        \includegraphics[width=\textwidth]{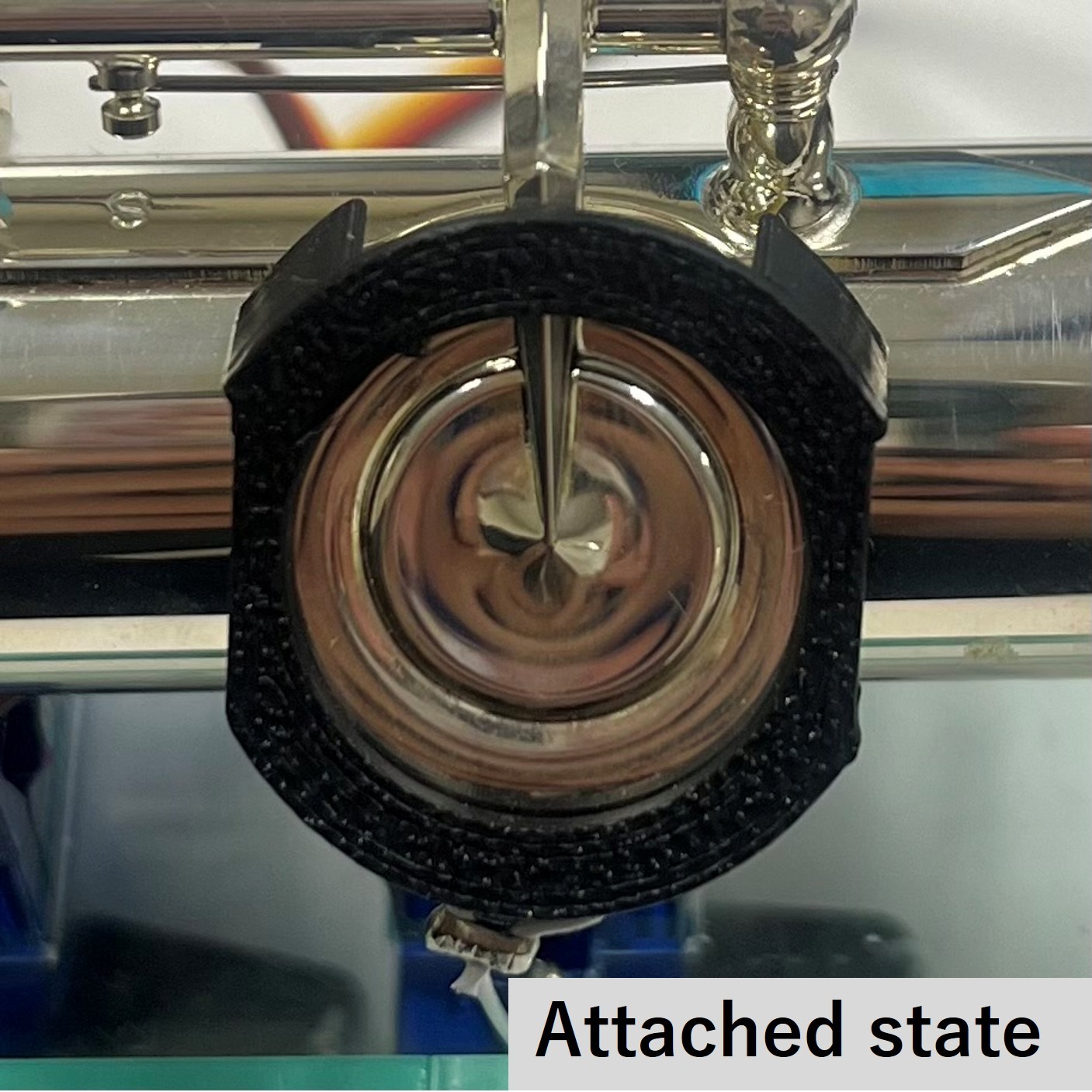}
        \vspace{-8pt}
        {\small (a)} 
    \end{minipage}
    \hspace{0.05\textwidth} 
    \begin{minipage}[b]{0.3\textwidth}
        \centering
        \includegraphics[width=\textwidth]{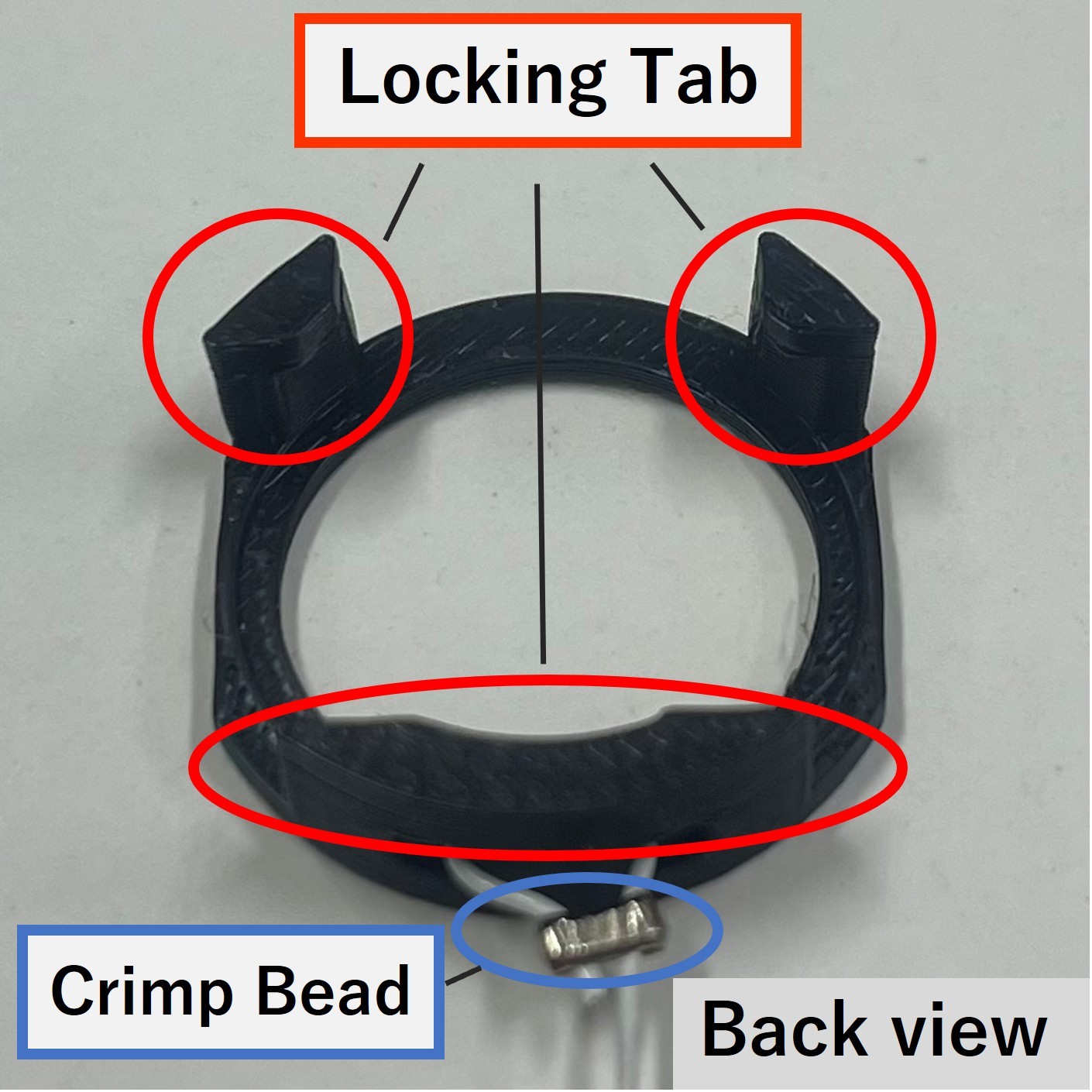}
        \vspace{-8pt}
        {\small (b)}
    \end{minipage}
    \vspace{5pt}
    \caption{Key cover structure. (a) Cover attached to key. 
    (b) Underside: three locking tabs for attachment; wire passes 
    through one tab and is fixed with a crimp bead.}
    \label{fig:keycover}
\end{figure}

Standard keys are fitted with removable key covers 
(Figure~\ref{fig:keycover}). 
The wire passes through a hole in one of three locking tabs on the underside of the cover and is fixed at the calibrated position using a crimp bead; the locking tabs hook onto the back of the key to prevent detachment during performance. 
For levers, which cannot accept key covers due to their structure and which open rather than close tone holes, wires are attached directly with sufficient slack for removal.

\paragraph{\underline{Rack-and-Pinion Method}}

\begin{figure}[tb]
    \centering
    \begin{minipage}[b]{0.38\textwidth}
        \centering
        \includegraphics[width=\textwidth]{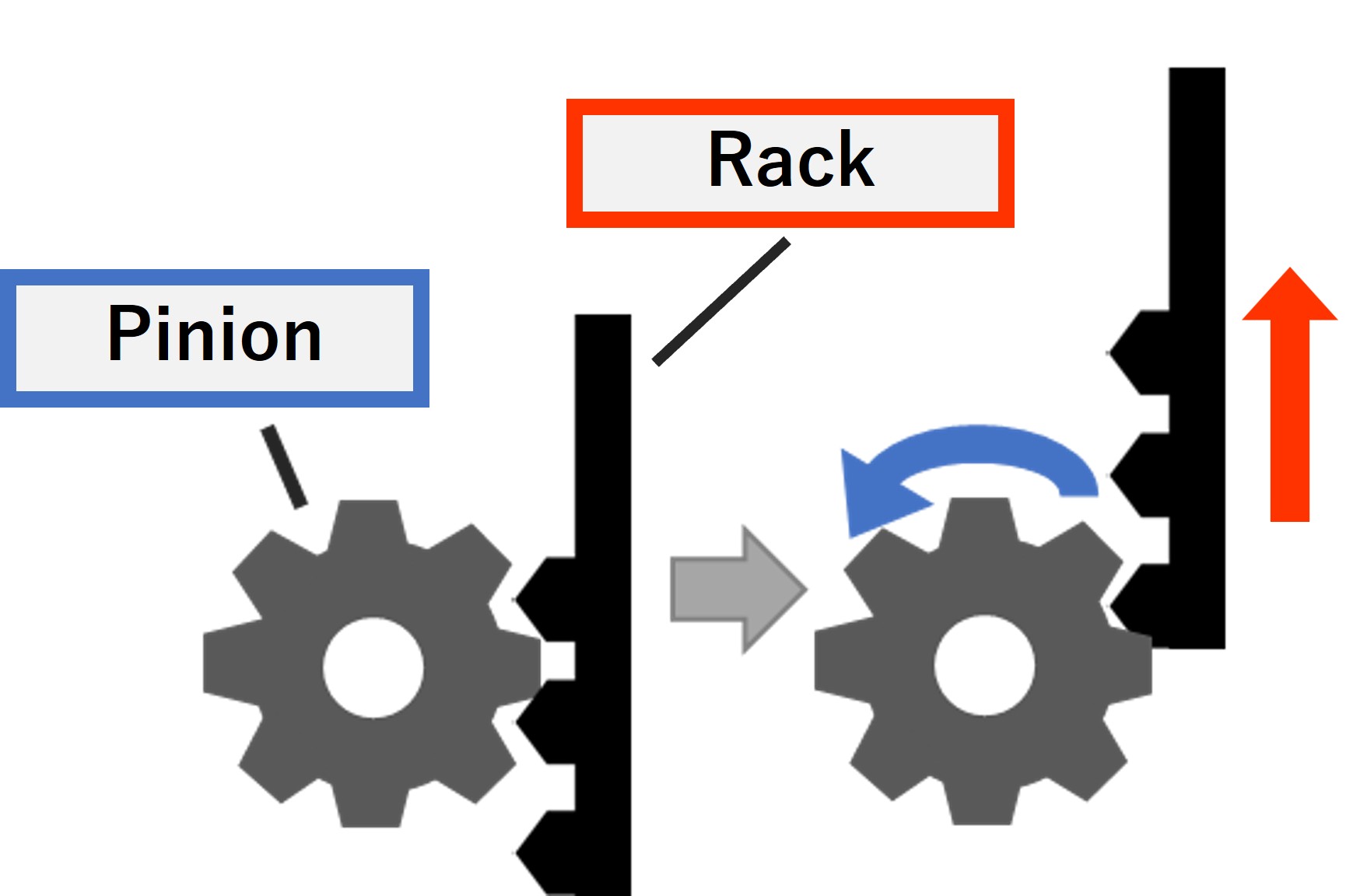}
        \vspace{-8pt} 
        {\small (a)} 
    \end{minipage}
    \hspace{0.05\textwidth} 
    \begin{minipage}[b]{0.38\textwidth}
        \centering
        \includegraphics[width=\textwidth]{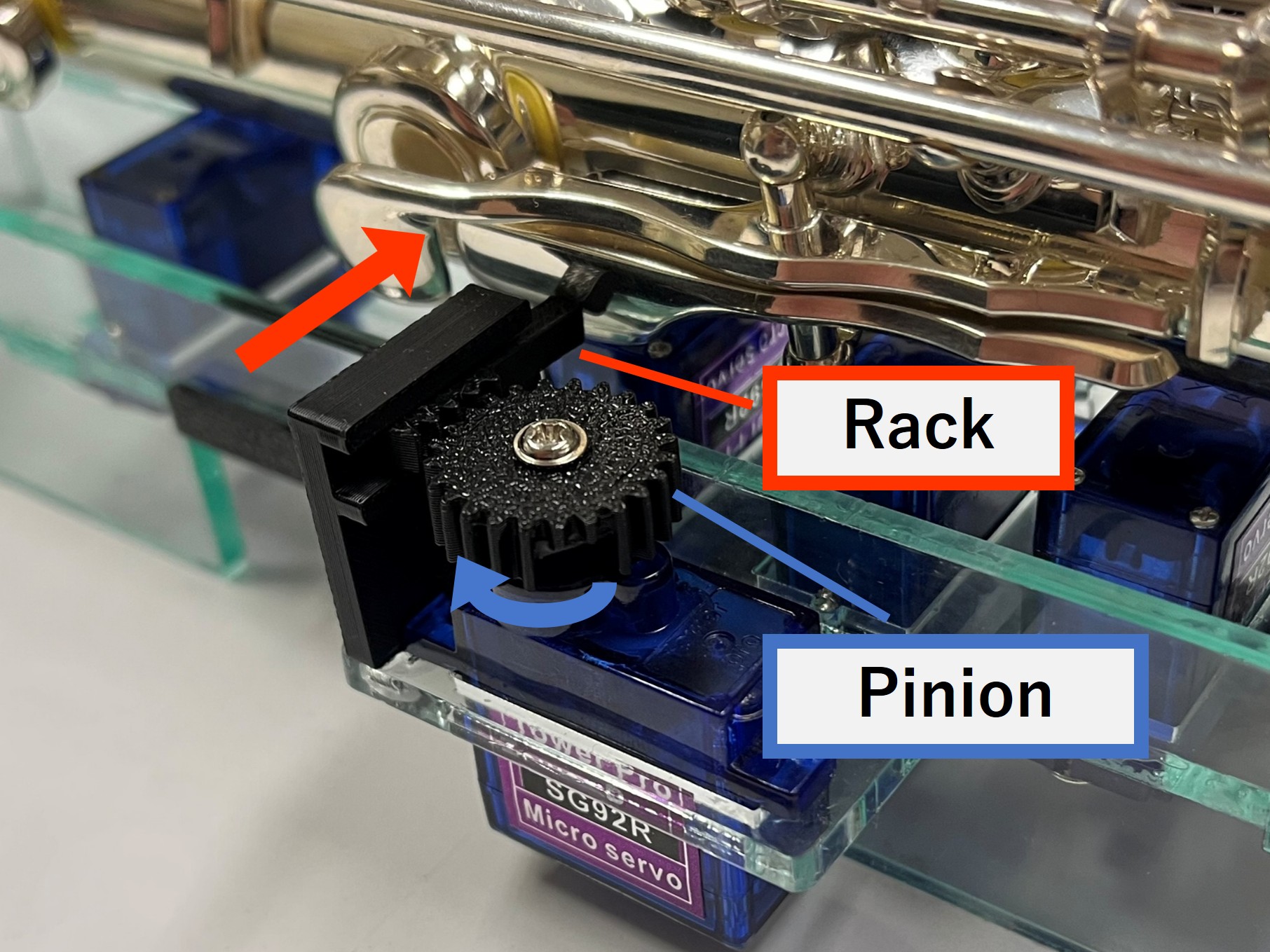}
        \vspace{-8pt}
        {\small (b)} 
    \end{minipage}
    \vspace{5pt}
    \caption{Rack-and-pinion mechanism for side-mounted keys. 
    (a) Schematic of the conversion of motor rotation to linear motion. 
    (b) Key actuation in use.}
    \label{fig:rack_pinion}
\end{figure}

Side-mounted keys are unlikely to be reliably actuated by wire pull because the actuation direction is not aligned with key travel.
For these keys, a rack-and-pinion mechanism converts motor rotation 
into linear motion (Figure~\ref{fig:rack_pinion}): pinion rotation 
drives the rack along the key travel axis, and the rack end presses 
the key directly. 
This method provides stable and repeatable closure of side-mounted 
tone holes.

\subsection{Jet Offset Assist Mechanism}
\label{subsec:jetoffset}

\subsubsection{Design Rationale}

\begin{figure}[tb]
    \centering 
    \includegraphics[width=0.7\linewidth]{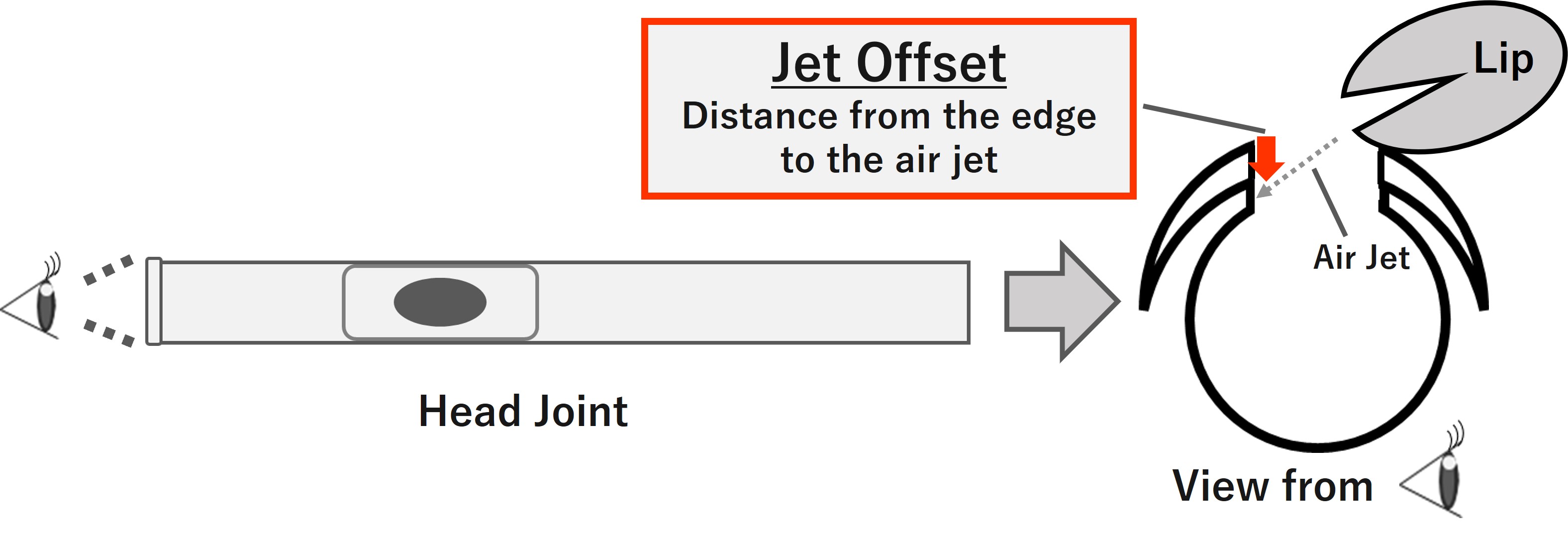}
    \caption{Jet offset: the displacement of the air jet center from the embouchure hole edge toward the inside of the flute (red arrow).}
    \label{fig:henshin}
\end{figure}

The jet offset is the distance by which the center of the air jet is 
displaced toward the inside of the flute from the edge of the embouchure 
hole (Figure~\ref{fig:henshin})~\cite{ando_book}.
The optimal jet offset for the low register differs from that of the middle and high registers, which share the same optimum~\cite{ando_paper1}. 
Because many low- and middle-register notes share identical fingerings, register is determined primarily by air jet velocity and jet offset. 
When the jet offset is set appropriately for the low register, the pitch does not easily transition to the middle register as velocity increases~\cite{ando_book}; this also permits greater dynamic range in the low register.
Experienced players adjust jet offset by modifying their embouchure; however, this control can be difficult for many beginners and may contribute to difficulties in producing low-register notes.

The proposed mechanism does not teach embouchure technique. 
Its purpose is to shift the jet offset mechanically toward a value appropriate for the low register during performance, reducing the acoustic difficulty of low-register playing without requiring embouchure control from the performer.

\subsubsection{Design Rationale for Head Joint Rotation}

In general, beginner players tend to find low-register notes more difficult to produce than middle-register notes.
Therefore, the proposed mechanism assumes that the jet offset of beginner players corresponds to a value appropriate for the middle or high register and shifts it toward the optimal value for the low register during low-register performance.

\begin{figure}[tb]
    \centering 
    \includegraphics[width=0.7\linewidth]{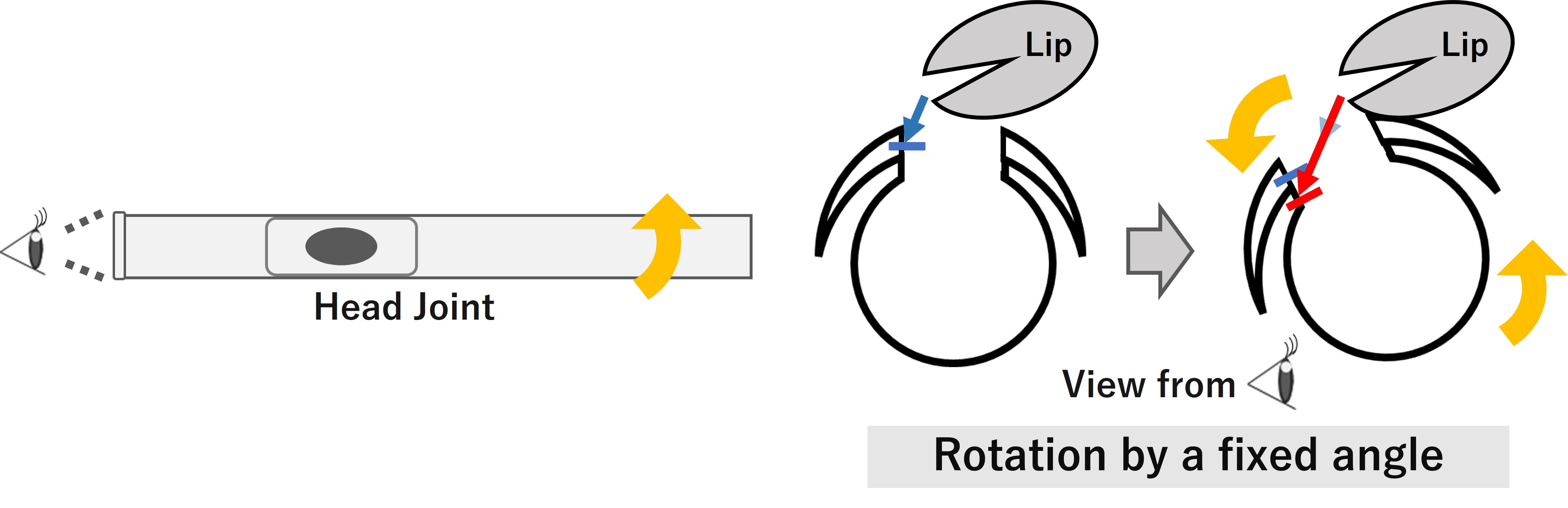}
    \caption{Principle of jet offset shift. The initial jet offset is represented by the blue line. Rotating the head joint by a predetermined angle shifts it to the red line.}
    \label{fig:henshin_idea}
\end{figure}

Rotating the head joint by a fixed angle changes the geometric relationship between the player's lip position and the embouchure hole, which is expected to shift the jet offset (Figure~\ref{fig:henshin_idea}).
This provides a mechanically simple, repeatable, and non-invasive means of altering jet offset without modifying the instrument or requiring any change in the performer's embouchure. 
The required rotation angle is determined through pre-performance calibration and is applied uniformly to all low-register notes.

\subsubsection{Mechanism Overview}

\begin{figure}[tb]
    \centering 
    \includegraphics[width=0.65\linewidth]{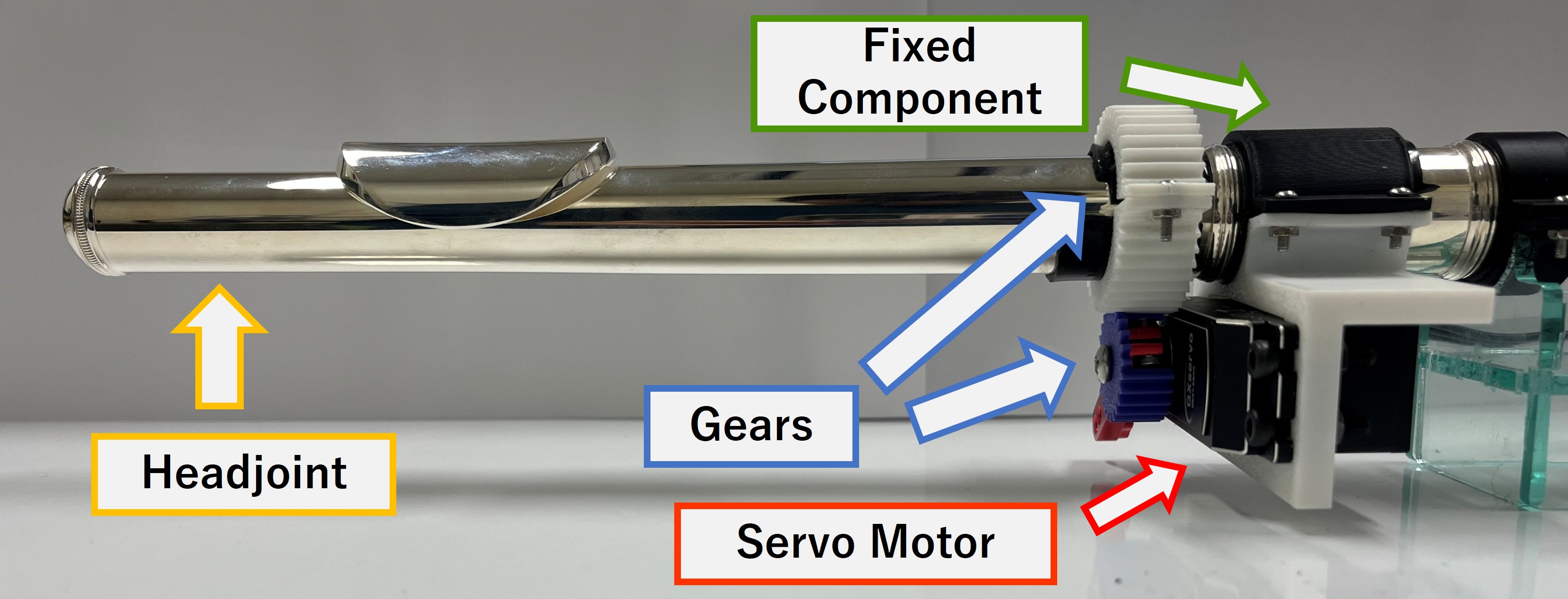}
    \caption{Jet offset assist mechanism mounted near the middle-to-head joint junction. The servo motor drives the head joint via meshing gears on the servo horn and joint body.}
    \label{fig:henshin_all}
\end{figure}

The mechanism is mounted near the junction of the head and middle joints (Figure~\ref{fig:henshin_all}), keeping the center of gravity close to the player's grip point to preserve balance during performance. 
All components are 3D-printed and fixed to the middle joint end with bolts and nuts; a silicone sheet between the attachment component and the flute prevents slippage without damaging the instrument.
The actuator is an X25 servo motor (GX Servo). 
Meshing gears on the servo horn and the head joint couple motor rotation directly to head joint rotation.

\subsection{Control System}
\label{subsec:control}

\begin{figure}[tb]
    \centering 
    \includegraphics[width=0.65\linewidth]{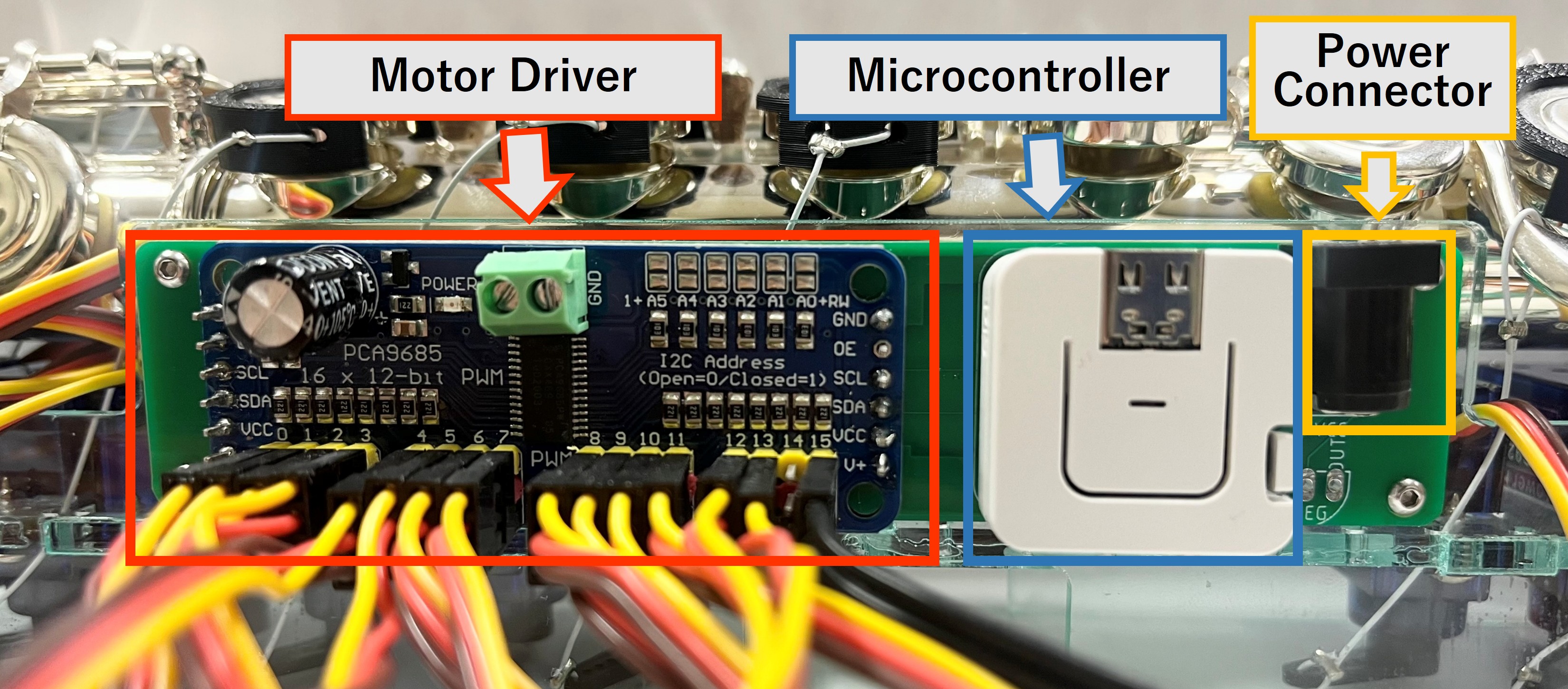}
    \caption{Configuration around the circuit board.From right to left, the power connector, microcontroller, and motor driver are mounted. The servo motors are controlled by these components.
    }
    \label{fig:board}
\end{figure}

Figure~\ref{fig:board} shows the control board configuration. 
The board carries an AtomS3-Lite microcontroller (M5Stack) and a PCA9685 PWM motor driver (NXP), powered via a regulated supply. 
All servo motors in both mechanisms are driven through this single board.

The signal path is as follows. A MIDI sequencer on a host PC transmits Note-On messages to the microcontroller. For each incoming note number, the microcontroller looks up the pre-calibrated servo positions assigned to the corresponding fingering and drives the motors accordingly. The jet offset assist mechanism is triggered automatically when the note number falls within the low-register range, rotating the head joint to the calibrated low-register angle; it returns to the initial position when a non-low-register note is received.

\section{Experiments and Results}
\label{sec:validation}

This section evaluates the mechanical feasibility of the two proposed mechanisms. For the automatic fingering mechanism, three experiments assess pitch accuracy across the full playing range, pitch accuracy during musical piece performance, and key/lever movement time. 
For the jet offset assist mechanism, two experiments assess whether the mechanism produces the intended change in jet offset during musical performance, and measure motor and head joint movement time. All evaluations target mechanical behavior and performance limits; user-centered evaluation is left as future work.


\subsection{Automatic Fingering Mechanism}
\label{subsec:val_fingering}

\subsubsection{Experimental Setup}

Experiments [A1] and [A2] were conducted in a soundproof room. 
The flute used was a YAMAHA YFL-412. Audio was recorded via a marantz Professional MPM1000 microphone through a YAMAHA MG10XUF mixer, with power supplied by a HANMATEK HM310 regulated supply.
Performance data for both experiments were recorded by a single flutist with seven years of playing experience.

For experiment [A3], key and motor movements were recorded using a CASIO EX-ZR1000 high-speed camera at 240~fps in the same equipment configuration.

\subsubsection{Experiment [A1]: Played Note Frequency Accuracy — Full Range}

\paragraph{\underline{Objective and Metric}}

This experiment verifies that the automatic fingering mechanism produces correct pitches across the full pitch range of the flute (C4–C7). 
The Harvest algorithm~\cite{Harvest} was used to estimate the played note frequency $f_0$ from recorded audio.
A detected $f_0$ within $\pm$50~cent of the target frequency is taken as confirmation that the intended pitch is produced. This threshold is based on the definition of one semitone as 100 cents; deviations within ±50 cents are therefore adopted here as a mechanical pass criterion rather than a standard of musical accuracy.
Spectrograms were additionally computed with a 12~ms window to resolve transient mechanical events at fingering transitions.

\paragraph{\underline{Procedure}}

\begin{figure}[tb]
    \centering
    \includegraphics[width=1\linewidth]{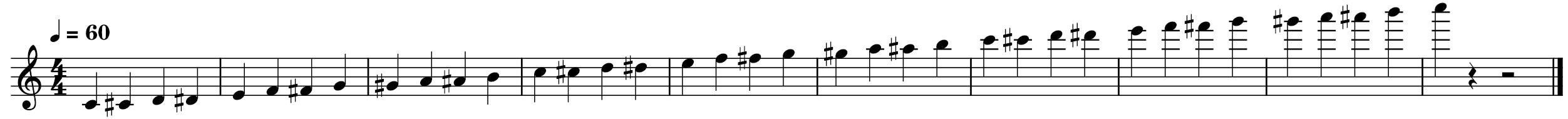}
    \caption{Score corresponding to the MIDI messages used in experiment [A1].
    The piece is a chromatic scale from the lowest note C4 to the highest note C7.}
    \label{fig:scale_score}
\end{figure}

MIDI messages corresponding to a chromatic scale from C4 to C7 (Figure~\ref{fig:scale_score}) were input to the mechanism.
The performer held the body of the mechanism and blew into the embouchure hole without touching any keys.

\paragraph{\underline{Results}}

\begin{figure}[tb]
    \begin{adjustwidth}{-\extralength}{0cm}
    \centering
    \begin{minipage}{0.48\linewidth}
        \centering
        \includegraphics[width=\linewidth]{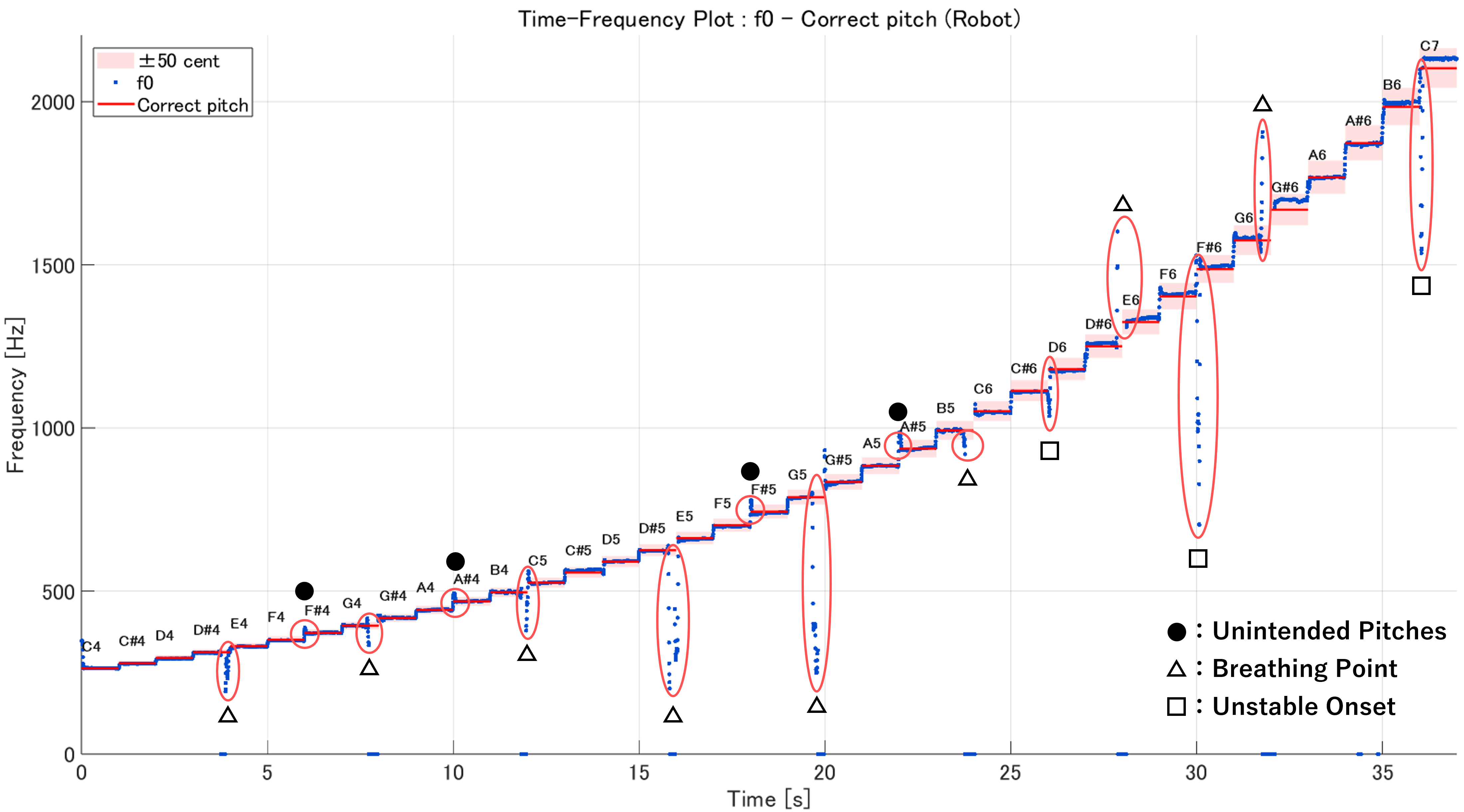}
        \vspace{2pt}
        {\small (a)}
    \end{minipage}
    \hfill
    \begin{minipage}{0.48\linewidth}
        \centering
        \includegraphics[width=\linewidth]{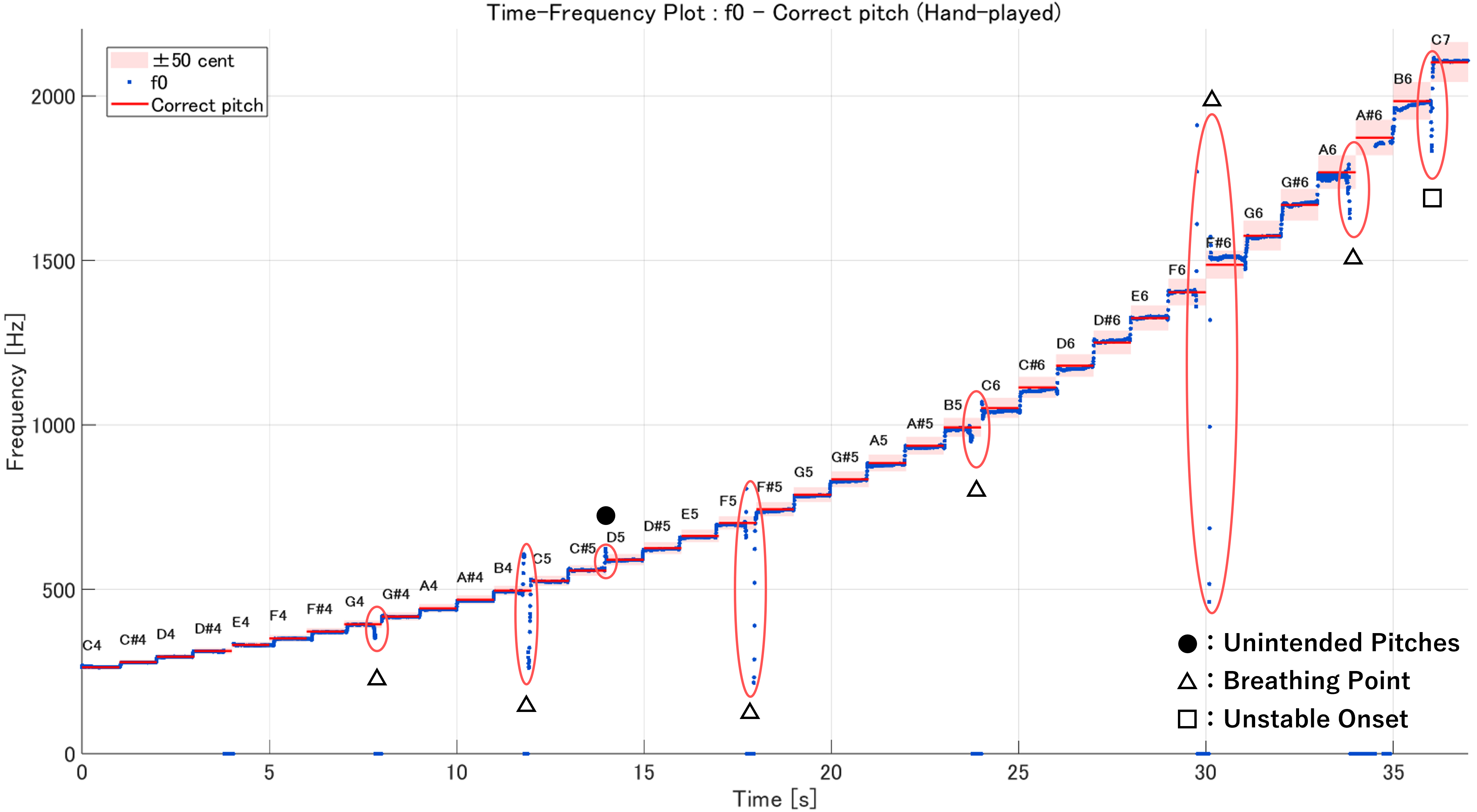}
        \vspace{2pt}
        {\small (b)}
    \end{minipage}
    \caption{$f_0$ estimation for chromatic scale C4--C7. 
    (a) Automatic fingering mechanism. 
    (b) Manual performance (reference). 
    Red line: target $f_0$; blue dots: estimated $f_0$; 
    light red band: $\pm$50~cent tolerance. 
    \raisebox{-0.25ex}{\Large $\bullet$}: unintended pitch; 
    $\triangle$: breath; $\square$: unstable onset.}
    \label{fig:scale_compare}
    \end{adjustwidth}
\end{figure}

Figure~\ref{fig:scale_compare}(a) shows the $f_0$ estimation results for the automatic fingering mechanism; Figure~\ref{fig:scale_compare}(b) shows manual performance for reference.
In both figures, the red line indicates the target $f_0$, blue dots the estimated $f_0$, and the light red band the $\pm$50~cent tolerance. 
The symbol \raisebox{-0.25ex}{\Large $\bullet$} marks unintended transient pitches, $\triangle$ marks breathing points, and $\square$ marks segments with unstable onsets and no clearly defined pitch.

The mechanism produces the intended pitch for all notes across the full range.
Unintended transient pitches occur at two fingering transitions (F--F$\sharp$ and A--A$\sharp$), which are attributed to the mechanism’s sensitivity to slight wire-induced timing differences between key presses.
Breath points and onset-unstable segments appear in both the robotic and manual performances, indicating these are related to airflow rather than mechanical error.

\begin{figure}[tb]
    \begin{adjustwidth}{-\extralength}{0cm}
    \centering
    \begin{minipage}{0.48\linewidth}
        \centering
        \includegraphics[width=\linewidth]{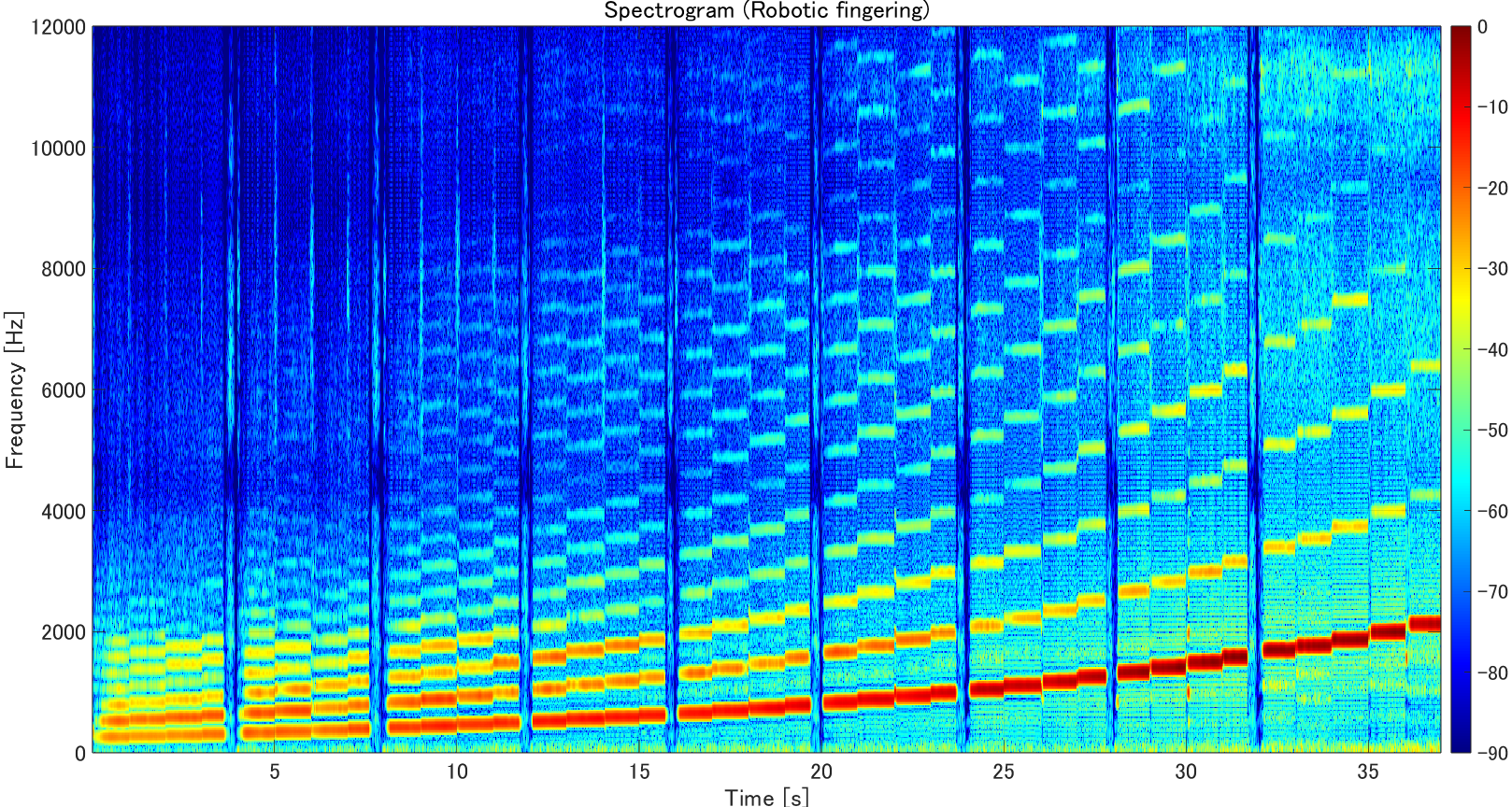}
        \vspace{2pt}
        {\small (a)}
    \end{minipage}
    \hfill
    \begin{minipage}{0.48\linewidth}
        \centering
        \includegraphics[width=\linewidth]{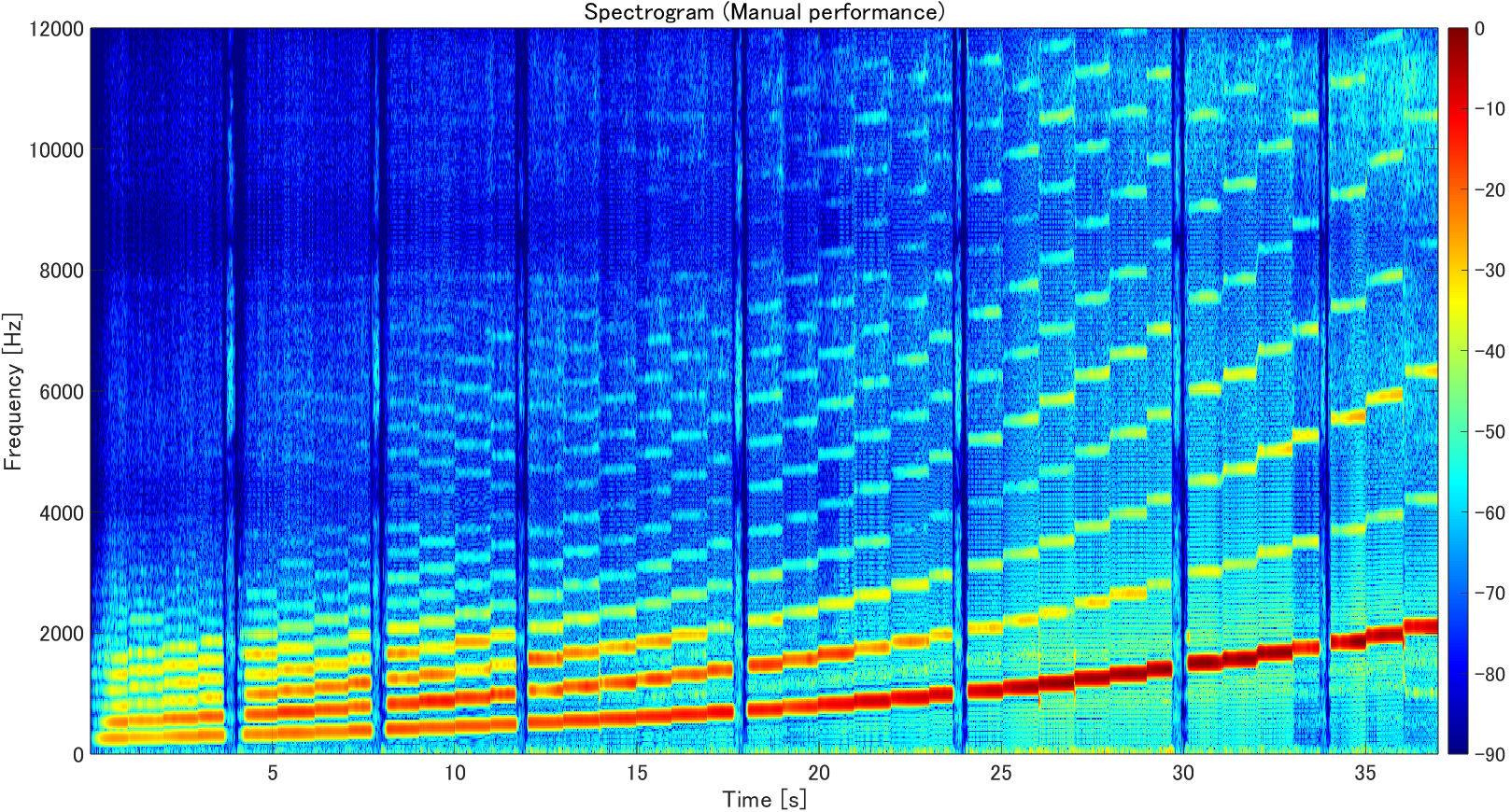}
        \vspace{2pt}
        {\small (b)}
    \end{minipage}
    \caption{Spectrograms of chromatic scale C4--C7. 
    (a) Automatic fingering mechanism: vertical streaks at fingering transitions indicate motor and key-click noise. (b) Manual performance (reference): no such components.}
    \label{fig:spe_scale_compare}
    \end{adjustwidth}
\end{figure}

Figure~\ref{fig:spe_scale_compare} shows spectrograms of the same performances.
Compared with manual performance, the automatic fingering mechanism exhibits short-duration broadband energy at each fingering transition, appearing as vertical streaks. 
These are likely attributable to motor actuation noise and key click noise rather than flute harmonics, given their temporal alignment with key-change events and their broadband rather than harmonic characteristics.

\subsubsection{Experiment [A2]: Played Note Frequency Accuracy — Musical Piece}

\paragraph{\underline{Objective and Metric}}

This experiment verifies that the mechanism produces a correct sequence of pitches during musical piece performance. 
The same $f_0$ estimation method and $\pm$50~cent criterion as in [A1] are applied. The detected pitch sequence is compared against the input MIDI messages as a reference for sequential note production.

\paragraph{\underline{Procedure}}

\begin{figure}[tb]
    \centering
    \includegraphics[width=1\linewidth]{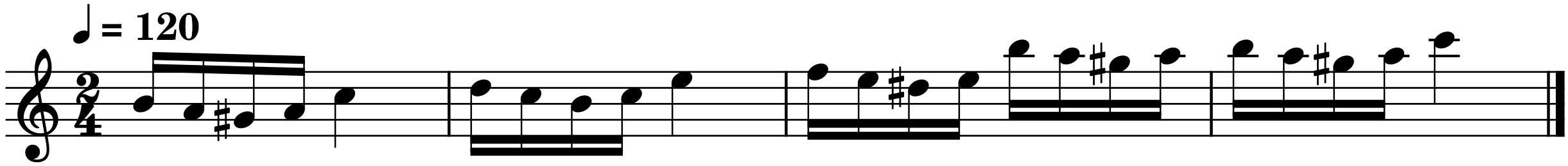}
    \caption{Score corresponding to the MIDI messages used in experiment [A2].
    The piece consists of the first four measures of Mozart's \textit{Turkish March}.}
    \label{fig:torco_score}
\end{figure}

MIDI messages corresponding to the first four measures of Mozart's Turkish March (Figure~\ref{fig:torco_score}) were input to the mechanism. 
The performer held the body and blew without touching any keys.

\paragraph{\underline{Results}}

\begin{figure}[tb]
    \begin{adjustwidth}{-\extralength}{0cm}
    \centering
    \begin{minipage}{0.48\linewidth}
        \centering
        \includegraphics[width=\linewidth]{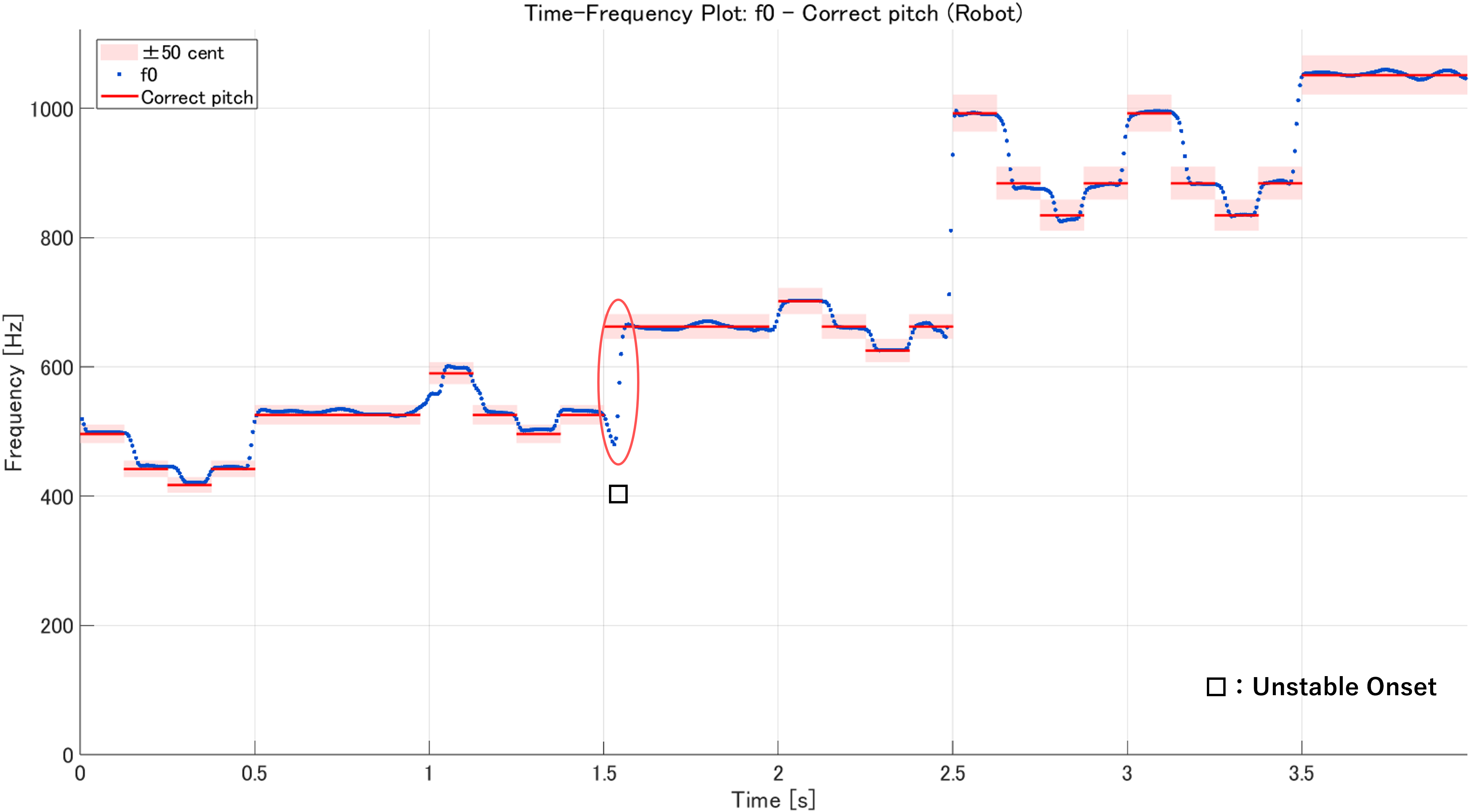}
        \vspace{2pt}
        {\small (a)}
    \end{minipage}
    \hfill
    \begin{minipage}{0.48\linewidth}
        \centering
        \includegraphics[width=\linewidth]{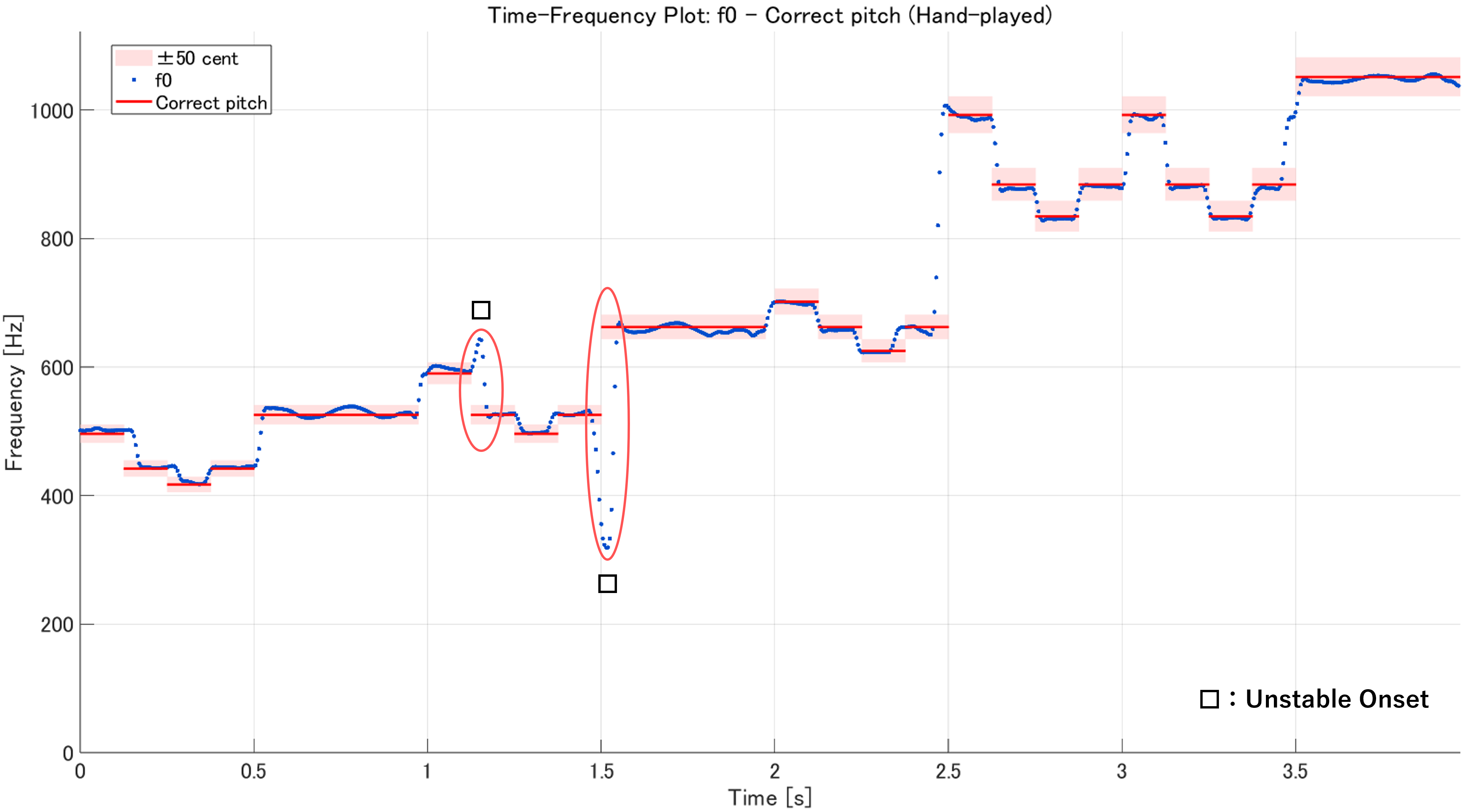}
        \vspace{2pt}
        {\small (b)}
    \end{minipage}
    \caption{$f_0$ estimation for Mozart's \textit{Turkish March}. 
    (a) Automatic fingering mechanism. 
    (b) Manual performance (reference). 
    Symbols as in Figure~\ref{fig:scale_compare}.}
    \label{fig:time-frequency_torco_compare}
    \end{adjustwidth}
\end{figure}

Figure~\ref{fig:time-frequency_torco_compare}(a) shows the estimated $f_0$ for the automatic fingering mechanism; Figure~\ref{fig:time-frequency_torco_compare}(b) shows the manual reference.
Notes are produced within the $\pm$50~cent tolerance throughout the piece. 
Onset-unstable segments ($\square$) occur in both conditions and are attributed to airflow variability rather than mechanical error. The time--frequency patterns of the two performances are broadly similar, confirming that the mechanism produces the intended note sequence.

\subsubsection{Experiment [A3]: Key and Motor Movement Time}

\paragraph{\underline{Objective and Metric}}

This experiment characterizes the actuation timing of each key and its driving motor to assess the mechanism's temporal performance and estimate the maximum supportable BPM. 
Motor movement time is defined as the interval from visually observed motor onset to cessation; key/lever movement time is defined as the interval from the onset of key closure to completion of the closing motion.

\paragraph{\underline{Procedure}}

Each key-closing action was recorded at 240~fps with a high-speed camera. Frames were counted from motion onset to cessation. 
Each action was repeated 10 times; mean duration and standard deviation were computed.

\paragraph{\underline{Results}}

Table~\ref{tab:key_times} summarizes motor and key/lever movement times across all keys. The C~key actuated by the motor corresponds to the dashed-line component in Figure~\ref{fig:flute}, not the solid-line C~key; this naming is used consistently throughout.

Motor movement time exceeds key/lever movement time for all wire-driven keys, reflecting a delay between motor actuation onset and the beginning of key travel, plus extra motor motion after the key reaches the fully closed position — both inherent to the wire method.
The B~key, driven by the rack-and-pinion  method,  shows the smallest motor-to-key time difference (48.75~ms vs.\ 45.42~ms).

Differences in movement times across key types are thought to result from variations in the required force—depending on whether a single key or multiple keys move in coordination—and differences in the spring characteristics of each key and lever.
These differences mean that key-closing timing is not perfectly uniform across all notes during multi-key transitions, and the transient pitch deviations observed in Experiment [A1] at the F--F$\sharp$ and A--A$\sharp$ transitions are consistent with this finding.

Using the longest measured key movement time (E key: 77.50~ms) as the limiting factor, the theoretical maximum BPM for each note value is shown in Table~\ref{tab:bpm}. 
At quarter-note resolution, the theoretical upper limit is 774~BPM, well above the fastest standard tempo markings (Presto/Prestissimo: approximately 170--200~BPM), confirming that the mechanism is not a tempo-limiting constraint for typical musical repertoire. 
These values represent theoretical upper bounds; in practice, the effective BPM is slightly lower due to the delay between motor start and effective key travel.

\begin{table}[tb]
\centering
\caption{Motor and key/lever movement times (mean $\pm$ SD over 
10 trials).}
\label{tab:key_times}
\begin{tabular}{lrr|rr}
\toprule
\textbf{Type} & 
\multicolumn{2}{c|}{\textbf{Motor}} & 
\multicolumn{2}{c}{\textbf{Key/Lever}} \\ 
\cmidrule(lr){2-3} \cmidrule(lr){4-5} 
 & \textbf{Avg.\ [ms]} & \textbf{SD~[ms]} 
 & \textbf{Avg.\ [ms]} & \textbf{SD~[ms]} \\
\midrule
LowC key          & 104.58 & 2.24 & 68.33 & 2.04 \\
C$\sharp$ key     & 104.17 & 1.86 & 57.50 & 3.12 \\
D$\sharp$ lever   &  57.50 & 1.67 & 29.58 & 2.24 \\
D key             &  91.67 & 1.86 & 70.42 & 1.25 \\
D$\sharp$ trill lever & 72.08 & 1.91 & 27.08 & 2.08 \\
E key             &  96.67 & 1.67 & 77.50 & 2.04 \\
D trill lever     &  65.00 & 2.04 & 31.25 & 2.08 \\
F key             &  99.58 & 1.25 & 68.75 & 2.08 \\
G$\sharp$ lever   &  79.58 & 1.25 & 60.00 & 2.04 \\
G key             & 101.25 & 1.91 & 77.08 & 2.08 \\
A key             & 103.75 & 1.25 & 76.25 & 1.91 \\
B$\flat$ key      &  99.17 & 1.67 & 67.92 & 1.91 \\
C key             &  97.92 & 2.08 & 60.42 & 2.08 \\
B key             &  48.75 & 1.91 & 45.42 & 1.25 \\
\bottomrule
\end{tabular}
\end{table}

\begin{table}[tb]
\centering
\caption{Theoretical maximum BPM of the automatic fingering mechanism, derived from the longest key movement time (E~key: 77.50~ms). One-measure duration shown for reference 
(4/4 time).}
\label{tab:bpm}
\begin{tabular}{lrr}
\toprule
\textbf{Note value} & \textbf{Max BPM} & 
\textbf{One measure [s]\footnotemark} \\
\midrule
Quarter note   & 774 & \multirow{3}{*}{0.31}  \\
Eighth note    & 387 &  \\
Sixteenth note & 194 &  \\
\bottomrule
\end{tabular}
\end{table}
\footnotetext{For reference, a typical fast piece at BPM200 corresponds to approximately 1.2~s per measure. }

\subsection{Jet Offset Assist Mechanism}
\label{subsec:val_jetoffset}

\subsubsection{Experimental Setup}

\begin{figure}[tb]
    \centering 
    \includegraphics[width=0.5\linewidth]{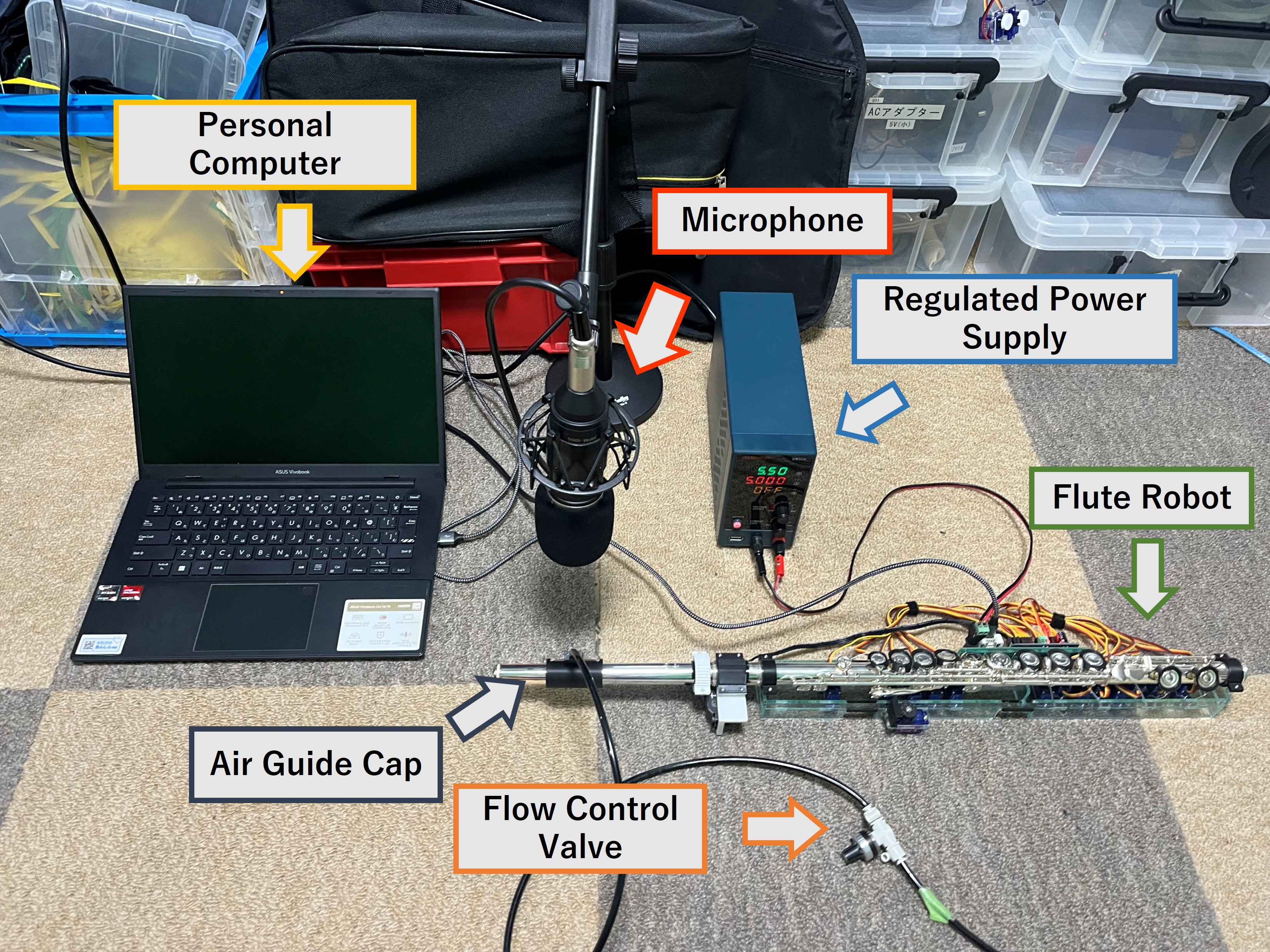}
    \caption{Experimental setup for jet offset assist mechanism evaluation. Soundproof room with air pump placed outside; flow rate controlled by needle valve; airflow direction fixed by 3D-printed cap.}
    \label{fig:env_henshin_ab}
\end{figure}

\begin{figure}[tb]
    \centering
    \begin{minipage}[b]{0.3\textwidth}
        \centering
        \includegraphics[width=\textwidth]{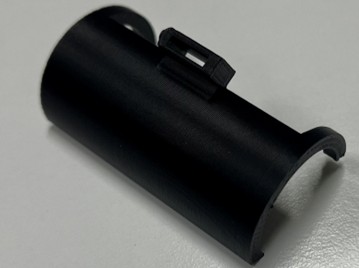}
        \vspace{-8pt} 
        {\small (a)} 
    \end{minipage}
    \hspace{0.03\textwidth} 
    \begin{minipage}[b]{0.3\textwidth}
        \centering
        \includegraphics[width=\textwidth]{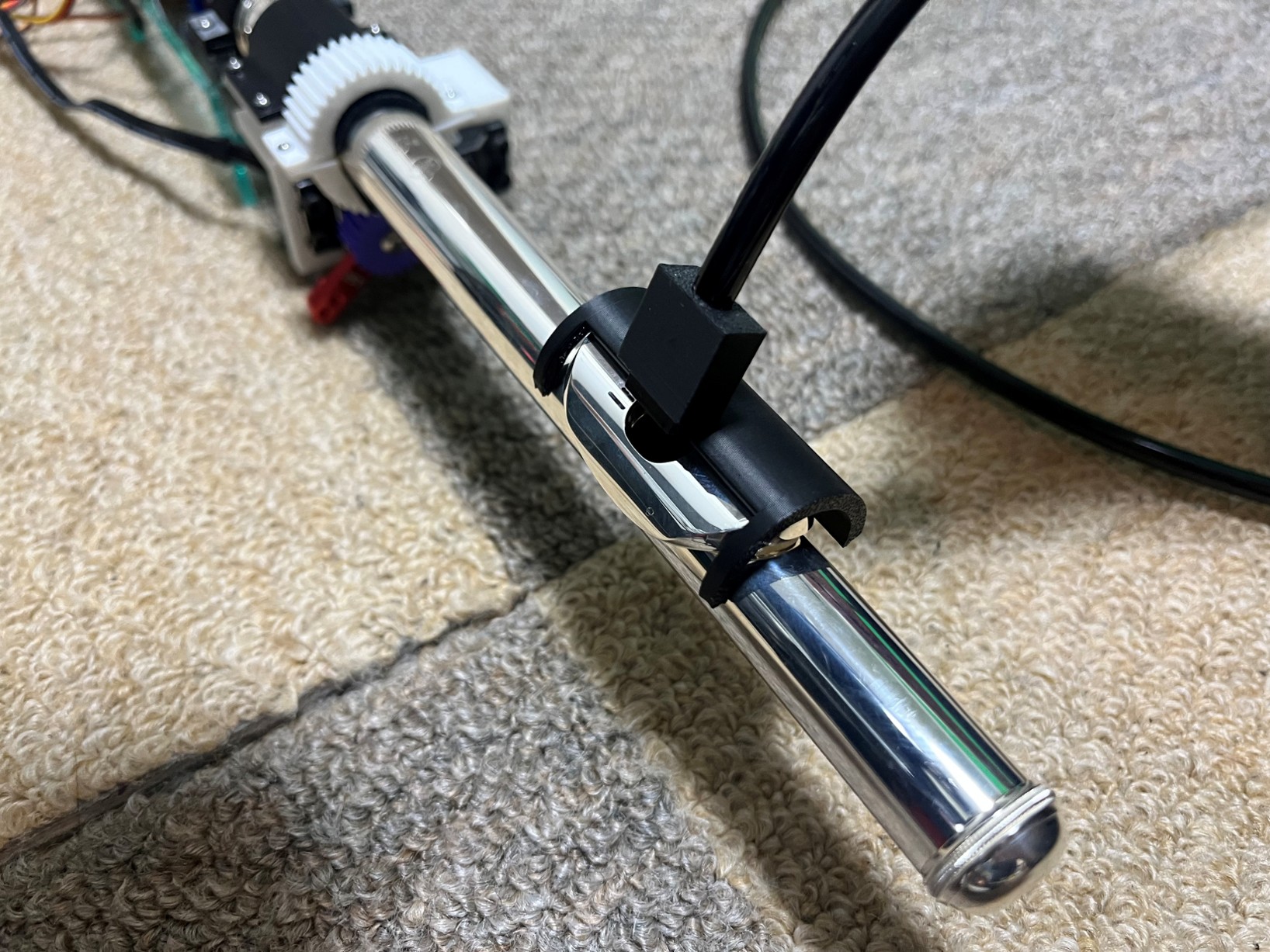}
        \vspace{-8pt}
        {\small (b)} 
    \end{minipage}
    \hspace{0.03\textwidth} 
    \begin{minipage}[b]{0.3\textwidth}
        \centering
        \includegraphics[width=\textwidth]{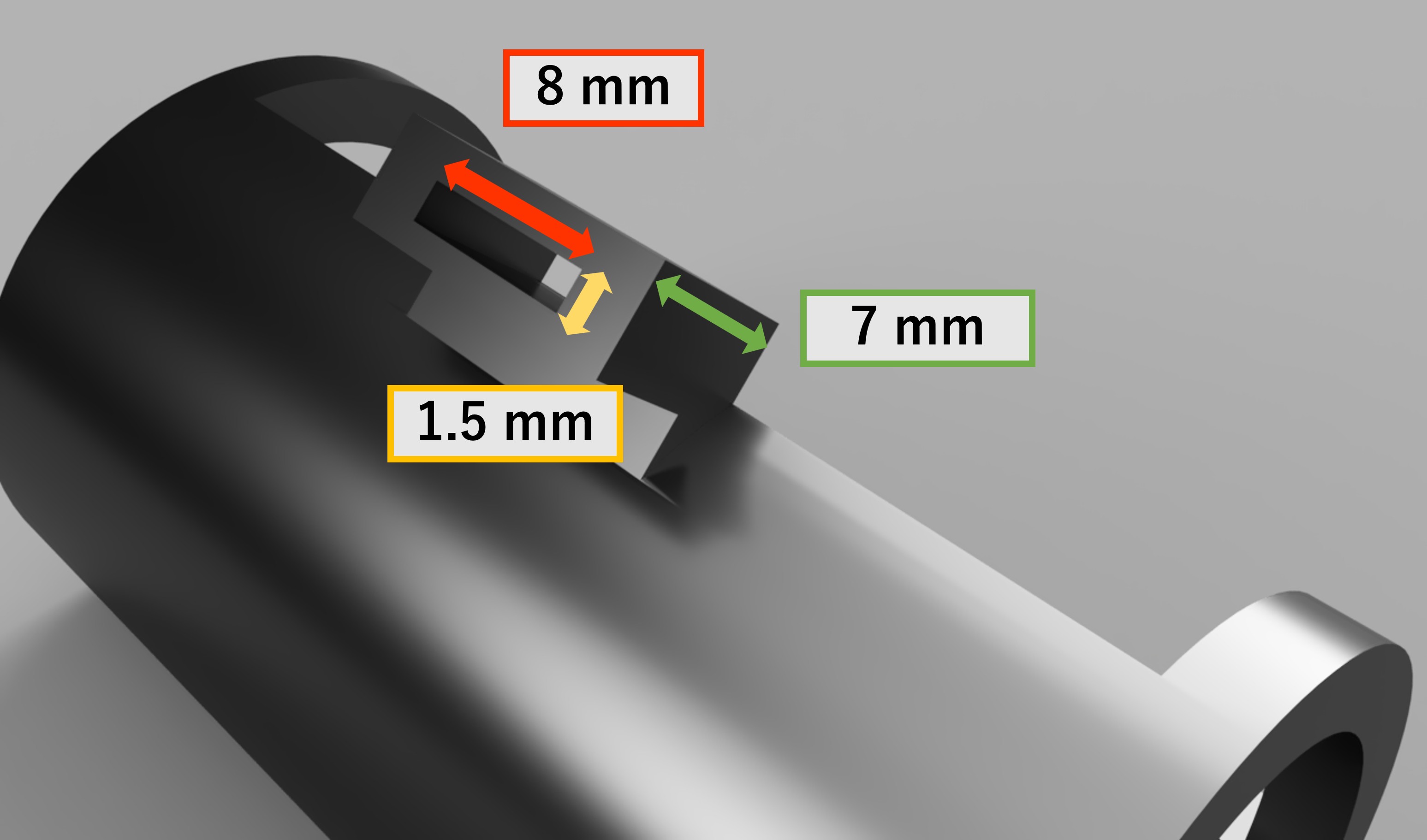} 
        \vspace{-8pt}
        {\small (c)}
    \end{minipage}
    \vspace{5pt}
    \caption{3D-printed airflow cap. (a) Cap. (b) Mounted on lip 
    plate. (c) Aperture dimensions. Rendered in Autodesk Fusion.}
    \label{fig:flow_cap}
\end{figure}

A constant-flow air supply was used in place of human blowing to eliminate airflow variability from the measurements. 
The experiment was conducted in a soundproof room (Figure~\ref{fig:env_henshin_ab}); the air pump (Yasunaga AP-60F) was placed outside to reduce operating noise from recordings, with flow rate controlled via a CKD DVL-S-06-H66-080 needle valve. 
Flute, microphone, and mixer specifications are as listed for experiments [A1]--[A2]. 
A 3D-printed cap (Figure~\ref{fig:flow_cap}) was fitted to the lip plate to keep the airflow direction relative to the embouchure hole constant.

For Experiment [B2], key movements were recorded at 240~fps with a CASIO EX-ZR1000 camera.

\begin{figure}[tb]
    \begin{adjustwidth}{-\extralength}{0cm}
    \centering
    \begin{minipage}{0.48\linewidth}
        \centering
        \includegraphics[width=\linewidth]{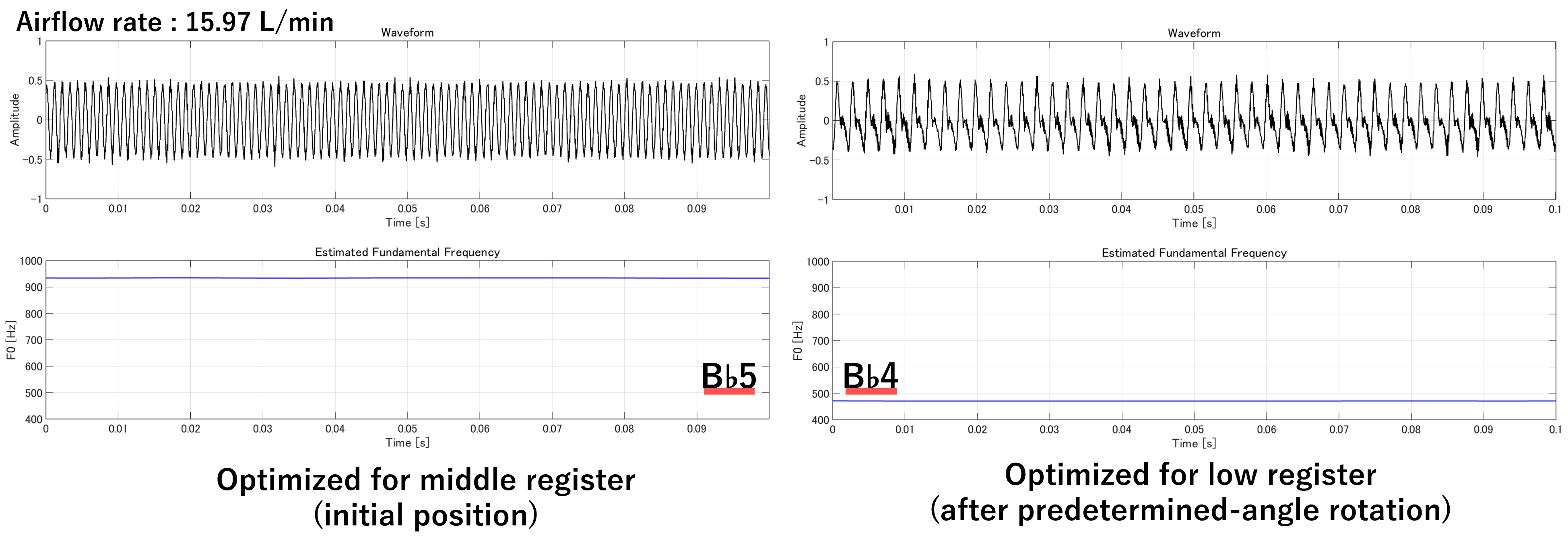}
        \vspace{2pt}
        {\small (a)}
    \end{minipage}
    \hfill
    \begin{minipage}{0.48\linewidth}
        \centering
        \includegraphics[width=\linewidth]{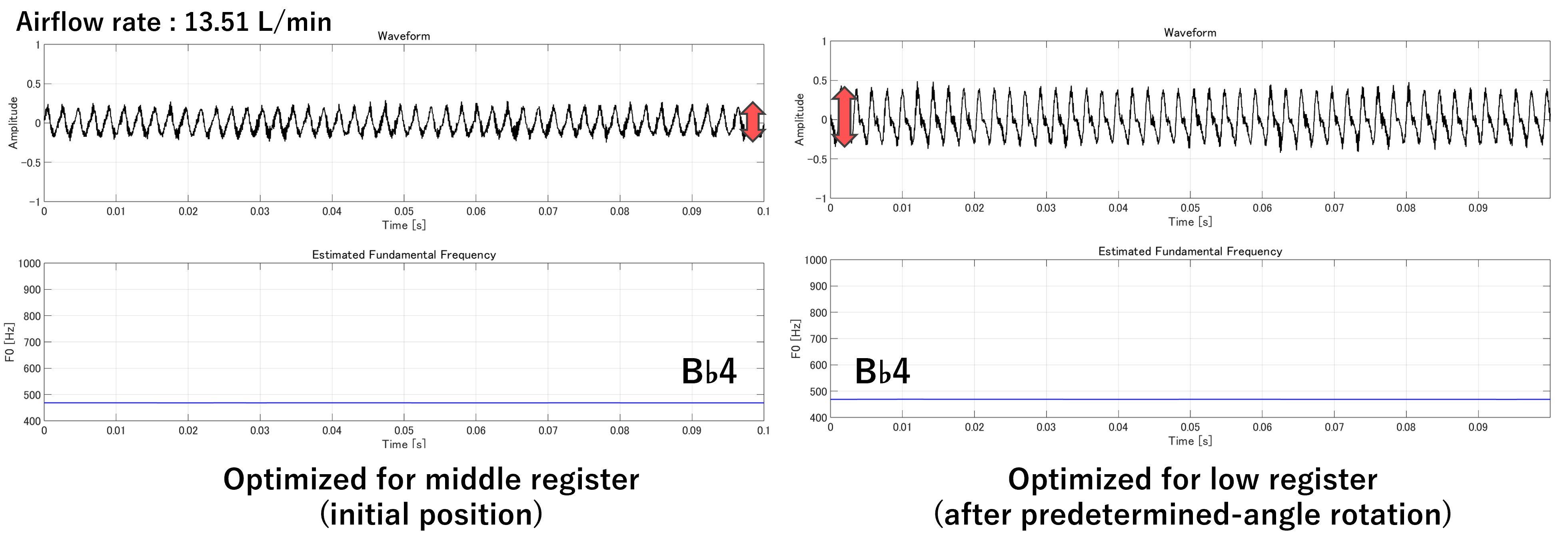}
        \vspace{2pt}
        {\small (b)}
    \end{minipage}
    \caption{Acoustic verification of head joint configurations 
    under identical airflow. 
    (a) 15.97~L/min: middle-register configuration produces 
    B$\flat$5; low-register configuration produces B$\flat$4. 
    (b) 13.51~L/min: both produce B$\flat$4, but the low-register 
    configuration shows larger amplitude.}
    \label{fig:position_flow}
    \end{adjustwidth}
\end{figure}

Prior to the main experiments, the two head joint configurations used throughout were verified acoustically under identical airflow conditions (Figure~\ref{fig:position_flow}):
\begin{itemize}
    \item \textit{Middle-register configuration} (initial position): 
    at 15.97~L/min airflow, this configuration produces B$\flat$5 
    (middle register).
    \item \textit{Low-register configuration} (after $22^\circ$ 
    rotation): at the same 15.97~L/min airflow, this configuration 
    produces B$\flat$4 (low register). At 13.51~L/min, both 
    configurations produce B$\flat$4, but the low-register 
    configuration yields larger amplitude.
\end{itemize}
These results confirmed that, under the conditions of this study, an approximately $22^\circ$ head joint rotation allows relatively stable production of low-register notes, and this angle was applied uniformly to all low-register notes in subsequent experiments.

\subsubsection{Experiment [B1]: Jet Offset Verification During Musical Performance}

\paragraph{\underline{Objective and Metric}}

This experiment verifies that the mechanism produces the intended change in jet offset during actual musical performance by comparing the harmonic structure of sounds produced with and without the mechanism. 
Changes in jet offset affect the relative amplitude of even and odd harmonics~\cite{ando_paper2, Onogi_2021}; accordingly, the difference in sound pressure level between the second and third harmonics ($\Delta\mathrm{SPL} = \mathrm{SPL}_2 - \mathrm{SPL}_3$) is used as a proxy for jet offset state. 
In the conditions of this study, higher-order harmonics were relatively small and susceptible to air pump noise; therefore, only the second and third harmonics were used.

\paragraph{\underline{Procedure}}

\begin{figure}[tb]
    \centering 
    \includegraphics[width=0.7\linewidth]{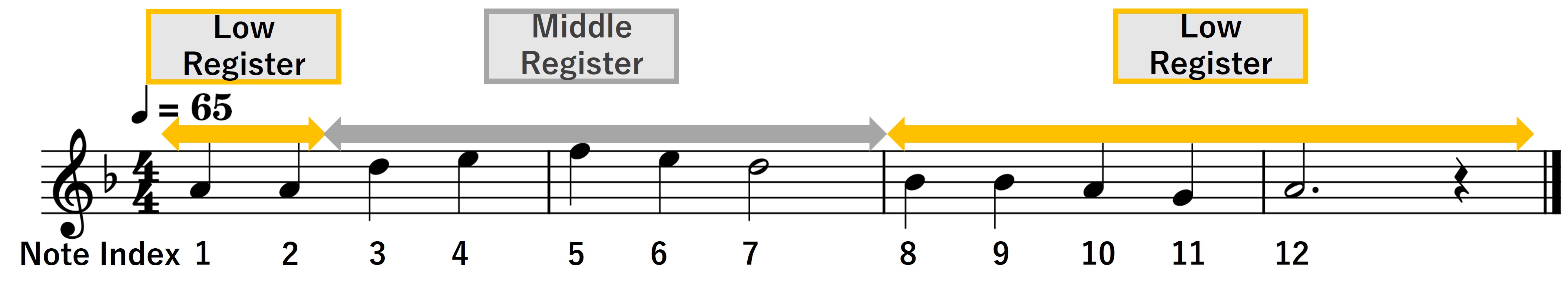}
    \caption{Score for Experiment [B1]: first four measures of 
    \textit{K{\=o}j{\=o} no Tsuki} (Rentar\=o Taki). 
    Yellow regions: low register (Note Indices 1--2, 8--12). 
    Gray region: middle register (Note Indices 3--7).}
    \label{fig:koujou_score}
\end{figure}

The first four measures of the melody from \textit{K{\=o}j{\=o} no Tsuki} (Rentar{\=o} Taki; Figure~\ref{fig:koujou_score}) were performed via MIDI input, with constant airflow from the air pump. 
This excerpt includes the low register in the first and last sections (Note Indices 1--2 and 8--12) and the middle register in the middle section (Note Indices 3--7). 
The experiment was repeated with and without the mechanism activated. For each note, the stable segment (excluding the first and last 100~ms) was extracted and a $2^{15}$-point FFT was applied. 
Harmonic peaks were identified by searching for maximum amplitude within $\pm$15~cent of twice and three times the target $f_0$.

\paragraph{\underline{Results}}

\begin{figure}[tb]
    \begin{adjustwidth}{-\extralength}{0cm}
    \centering 
    \includegraphics[width=0.7\linewidth]{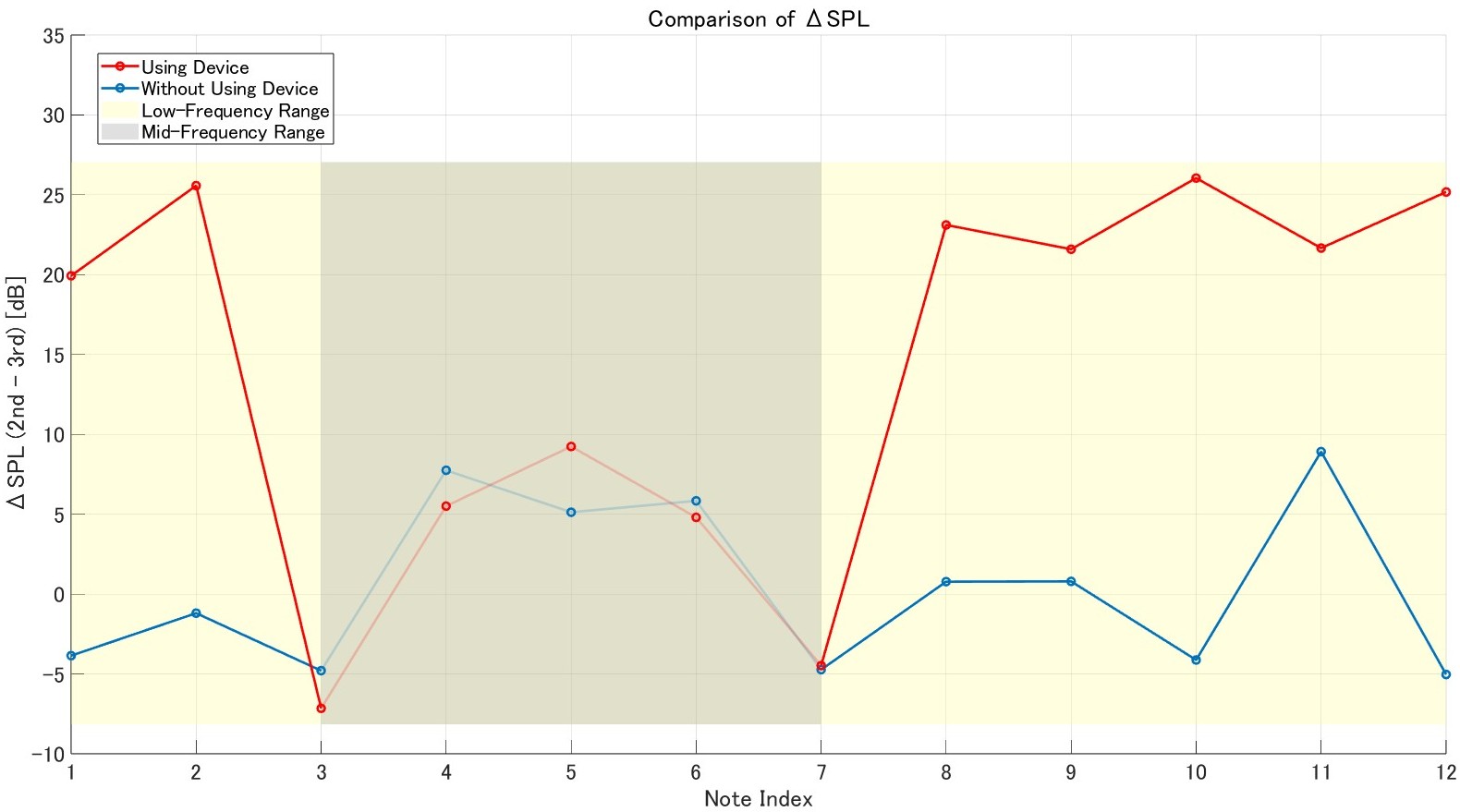}
    \caption{$\Delta$SPL (2nd harmonic [dB] $-$ 3rd harmonic [dB]) 
    per note, with and without the jet offset assist mechanism. 
    Red: mechanism active. Blue: mechanism inactive. 
    Yellow regions: low register. Gray region: middle register.}
    \label{fig:spl}
    \end{adjustwidth}
\end{figure}

Figure~\ref{fig:spl} shows $\Delta$SPL per note for both conditions.
When the mechanism was active (red), $\Delta$SPL increased for all low-register notes relative to the inactive condition (blue). 
No consistent difference was observed in the middle-register notes (Note Indices 3--7), as expected, since the mechanism does not actuate outside the low-register range.

According to Ando~\cite{ando_book}, $\Delta$SPL is small when jet offset is suited to the middle register and increases as the jet offset approaches the low-register optimum. 
The observed increase in $\Delta$SPL for all low-register notes is therefore consistent with the mechanism shifting the jet offset toward the low-register optimum during performance, confirming that it operates as intended.

\subsubsection{Experiment [B2]: Head Joint Movement Time}

\paragraph{\underline{Objective and Metric}}

This experiment characterizes the actuation timing of the jet offset assist mechanism to assess its suitability for musical performance tempos. Motor movement time and head joint rotation time are measured separately to identify the mechanical delay between motor actuation and effective joint movement.

\paragraph{\underline{Procedure}}

Head joint rotation to $22^\circ$ was recorded at 240~fps. 
Frames were counted from motion onset to cessation for both motor and head joint, repeated over 10 trials.

\paragraph{\underline{Results}}

Table~\ref{tab:motion_time} shows the results. 
Motor movement required 73.33~ms; head joint rotation completed in 40.00~ms. 
The 33~ms difference reflects the delay between motor actuation onset and the start of gear-coupled head joint motion, plus residual motor travel after joint completion.

Using the head joint rotation time of 40.00~ms as the limiting factor, the theoretical maximum BPM is shown in Table~\ref{tab:bpm_henshin}. 
At quarter-note resolution the upper limit is 1500~BPM, well above practical performance tempos, confirming that the mechanism does not impose a tempo constraint.

\begin{table}[tb]
    \centering
    \caption{Motor and head joint movement times (mean $\pm$ SD over 10 trials).}
    \label{tab:motion_time}
    \begin{tabular}{lrr}
        \toprule
        \textbf{Action} & \textbf{Duration [ms]} & \textbf{SD~[ms]}\\
        \midrule
        Motor movement      & 73.33 & 2.04 \\ 
        Head joint movement & 40.00 & 2.04 \\  
        \bottomrule
    \end{tabular}
\end{table}

\begin{table}[tb]
\centering
\caption{Theoretical maximum BPM of the jet offset assist mechanism, derived from head joint rotation time (40.00~ms). One-measure duration shown for reference (4/4 time).}
\label{tab:bpm_henshin}
\begin{tabular}{lrr}
\toprule
\textbf{Note value} & \textbf{Max BPM} & 
\textbf{One measure [s]} \\
\midrule
Quarter note   & 1500 & \multirow{3}{*}{0.16} \\
Eighth note    &  750 &  \\
Sixteenth note &  375 &  \\
\bottomrule
\end{tabular}
\end{table}

\subsection{Discussion}
\label{subsec:discussion}

\subsubsection{Automatic Fingering Mechanism}

Experiments [A1] and [A2] confirm that the mechanism produces the intended pitches across the full range of the flute and throughout a musical piece under controlled conditions. 
Transient pitch deviations at two specific fingering transitions (F--F$\sharp$ and A--A$\sharp$) are consistent with wire-induced timing differences when multiple keys sharing a fingering are actuated simultaneously.
Addressing this through optimized key-press sequencing or tighter wire tensioning is a priority for future mechanical refinement.

Motor actuation noise, visible as broadband frequency components in the spectrograms of Experiment [A1], is the most acoustically prominent characteristic distinguishing robotic from manual performance. 
The movement time data from Experiment [A3] confirm that timing irregularities across keys are present but small (SD $\leq$ 3.12~ms for key/lever movement), and BPM estimates 
indicate the mechanism operates comfortably within the tempo range of typical musical repertoire. 
Reducing motor actuation noise is identified as the primary mechanical design priority for future iterations.

\subsubsection{Jet Offset Assist Mechanism}

Experiment [B1] confirms that the mechanism produces a consistent increase in $\Delta$SPL across all low-register notes when activated, while leaving middle-register notes unaffected. 
This result is consistent with the mechanism shifting the jet offset toward a state suitable for the low register.
The evaluation was conducted under controlled airflow conditions; whether the same effect holds under variable human blowing requires future investigation.

Experiment [B2] confirms that head joint rotation completes within 40~ms, well within the timing constraints of practical musical performance. 
The 33~ms motor-to-joint delay is inherent to the gear-coupled drive and should be accounted for in MIDI trigger timing during system calibration.

Both mechanisms are evaluated here under controlled mechanical conditions. Evaluation of usability with beginner players, and assessment of whether the jet offset assist mechanism meaningfully reduces the difficulty of low-register note production for human performers, are left as directions for future work.


\section{Conclusion}
\label{sec:conclusion}

This study presented a semi-automatic flute robot comprising two independently operable mechanisms mounted on a standard covered-key flute: an automatic fingering mechanism and a jet offset assist mechanism.

The automatic fingering mechanism actuates all fourteen keys via servo motors in response to MIDI input, using wire-based and rack-and-pinion actuation depending on key geometry. 
Fundamental frequency estimation across the full chromatic range (C4--C7) and during musical piece performance confirmed that the mechanism produces intended pitches within a $\pm$50~cent criterion throughout. 
Key/lever movement times across all keys remain within 77.50~ms, yielding a theoretical tempo capacity well above standard performance ranges. 

Motor actuation noise, identified as broadband frequency transients at fingering transitions, is the primary acoustic limitation of the current design.

The jet offset assist mechanism rotates the head joint by a calibrated angle of approximately $22^\circ$ during low-register passages, shifting the jet offset toward the low-register optimum without modifying the instrument or requiring embouchure adjustment from the performer.
Harmonic analysis ($\Delta\mathrm{SPL} = \mathrm{SPL}_2 - \mathrm{SPL}_3$) confirmed a consistent increase in $\Delta$SPL across all low-register notes when the mechanism was active, consistent with the intended jet offset shift. 
Head joint rotation completes within 40.00~ms, supporting performance at standard musical tempos.

Together, these results confirm the mechanical feasibility of both mechanisms under controlled conditions and establish a basis for further development.

\section{Limitations and Future Work}
\label{sec:futurework}

Several limitations of the present study should be noted, along with directions for future research.

\paragraph{\underline{Key actuation timing}}

Small differences in key movement timing when multiple motors operate simultaneously can cause temporary pitch deviations, as observed at the F--F\# and A--A\# transitions in Experiment~[A1].
The wire-based actuation method introduces a delay between motor onset and key travel, and this delay varies across keys (27.08--77.50\,ms).
To achieve reliable transitions, the key-press sequence could be optimized and possibly redesigned into a safe fingering pathway, as proposed for safe transitions by Almeida et al.~\cite{Almeida_2009}.

\paragraph{\underline{Motor drive noise}}

Motor actuation noise, identified as broadband frequency transients at fingering transitions, is the primary acoustic characteristic distinguishing the proposed system from manual performance.
Reducing this noise is considered the most critical design priority for future iterations.
Possible approaches include mechanical enclosure of drive components or replacement with lower-noise actuators.

\paragraph{\underline{Future Work}}

Two primary directions are identified for future development.
First, motor drive noise reduction should be pursued through mechanical enclosure of drive components or replacement with lower-noise actuators, as this is the dominant factor affecting the perceptual naturalness of the system.
Second, the jet offset assist mechanism was evaluated using a constant-flow air pump; whether the mechanism produces equivalent jet offset changes under variable human blowing remains to be verified. 

\vspace{6pt} 

\begin{adjustwidth}{-\extralength}{0cm}
\end{adjustwidth}
\end{document}